\documentclass[prd,english,preprintnumbers,amsmath,amssymb,nofootinbib,superscriptaddress,twocolumn]{revtex4-1}

\pdfoutput=1

\usepackage{mathtools}
\usepackage{amsfonts}
\usepackage{mathrsfs}
\usepackage{bbm}
\usepackage{slashed}

\usepackage{graphicx}
\usepackage[dvipsnames]{xcolor}
\usepackage{array}

\usepackage{xspace}
\usepackage{siunitx}
\usepackage{hyperref}

\usepackage{xifthen}
\usepackage{dsfont}
\usepackage{appendix}
\usepackage{enumitem}
\usepackage{booktabs}
\usepackage{units}
\usepackage{bigints}

\usepackage[latin1]{inputenc}
\usepackage[dvipsnames]{xcolor}
\usepackage{array}

\setkeys{Gin}{width=0.48\textwidth}

\graphicspath{
	{./}
}

\def\0#1#2{\frac{#1}{#2}}

\def\s0#1#2{\mbox{\small{$ \frac{#1}{#2} $}}}

\newcommand{\I}{\mathrm{i}}
\newcommand{\be}{\begin{eqnarray}}
\newcommand{\ee}{\end{eqnarray}}
\newcommand{\del}{\partial}

\newcommand{\nn}{\nonumber }

\newcommand{\tr}{\text{tr}}

\newcommand{\beq}{\begin{equation}}
\newcommand{\eeq}{\end{equation}}
\newcommand{\bea}{\begin{eqnarray}}
\newcommand{\eea}{\end{eqnarray}}

\newcommand{\diff}{\text{d}}
\newcommand{\muc}{\mu_{\text{c}}}

\def\0#1#2{\frac{#1}{#2}}

\hypersetup{
	colorlinks,
	linkcolor={red!75!black},
	citecolor={blue!75!black},
	urlcolor={blue!75!black},
	pdftitle={Renormalization Group Flows for Dense Relativistic Matter},
	pdfauthor={Braun, D\"ornfeld, Schallmo, T\"opfel},
	pdfkeywords={Renormalization group} {Dense Matter},
	bookmarksopen=true,
	bookmarksopenlevel=2,
	bookmarksnumbered=true
}

\usepackage{babel}
\makeatother
\begin{document}

\title{Renormalization Group Studies of Dense Relativistic Systems}

\author{Jens Braun}
\affiliation{Institut f\"ur Kernphysik, Technische Universit\"at Darmstadt, 
D-64289 Darmstadt, Germany}
\affiliation{ExtreMe Matter Institute EMMI, GSI, Planckstra{\ss}e 1, D-64291 Darmstadt, Germany}
\author{Timon D\"ornfeld} 
\affiliation{Institut f\"ur Kernphysik, Technische Universit\"at Darmstadt, 
D-64289 Darmstadt, Germany}
\author{Benedikt Schallmo} 
\affiliation{Institut f\"ur Kernphysik, Technische Universit\"at Darmstadt, 
D-64289 Darmstadt, Germany}
\author{Sebastian T\"opfel} 
\affiliation{Institut f\"ur Kernphysik, Technische Universit\"at Darmstadt, 
D-64289 Darmstadt, Germany}

\begin{abstract}
 Dense relativistic matter has attracted a lot of attention over many decades now, with a focus 
 on an understanding of the phase structure and thermodynamics of dense strong-interaction matter. 
 The analysis of dense strong-interaction matter is complicated by the fact that the system is expected 
 to undergo a transition from a regime governed by spontaneous chiral symmetry breaking at low densities to a regime 
 governed by the presence of a Cooper instability at intermediate and high densities. 
 Renormalization group (RG) approaches have played and still play a prominent role in studies of 
 dense matter in general. In the present work, we study RG
 flows of dense relativistic systems in the presence of a Cooper instability and analyze 
the role of the Silver-Blaze 
 property. In particular, we critically assess how to apply the derivative expansion to study dense-matter systems in a systematic fashion.
 This also involves a detailed discussion of regularization schemes. Guided by these formal developments, 
 we introduce a new class of regulator functions for functional RG studies which is suitable to deal 
 with the presence of a Cooper instability in relativistic theories. We close by demonstrating its application with the aid of a simple 
 quark-diquark model. 

\end{abstract}
\maketitle
%%%%%%%%%%%%%%%%%%%%%%%%%%%%%%%%%%%%%%%%%%%%%%%%%%%%%%%%%%
% 

%
\section{Introduction}
A quantitative understanding of relativistic fermions in a dense environment is of great importance 
for many research fields, ranging from condensed-matter physics over nuclear physics to high-energy physics. With respect 
to strong-interaction matter, the potential existence of a color-superconducting ground state 
at supranuclear densities put forward in the 1970s has 
inspired uncounted studies at the interface of nuclear physics and astrophysics (see Ref.~\cite{Bailin:1983bm} for an early review).
The interest in this state of matter 
even received a significant boost as a series of 
seminal works in the late 1990s~\cite{Alford:1997zt,Rapp:1997zu,Son:1998uk,Schafer:1998na,Berges:1998rc,Schafer:1999jg,Pisarski:1999tv,Pisarski:1999bf,Brown:1999aq,Hong:1999fh,Hsu:1999mp,Evans:1999at} 
suggested the emergence of 
a rich plethora of symmetry-breaking patterns at high densities together with the formation of sizeable pairing 
gaps of~$\sim 100\,\text{MeV}$ giving rise to large phase transition temperatures, 
see Refs.~\cite{Rajagopal:2000wf,Alford:2001dt,Buballa:2003qv,Shovkovy:2004me,Alford:2007xm,Fukushima:2010bq,Fukushima:2011jc,Anglani:2013gfu,Schmitt:2014eka} for reviews.
For the analysis of these symmetry-breaking patterns and the role of the Cooper instability associated with the formation of a pairing gap, 
renormalization group (RG) approaches have played and still play
a very important role~\cite{Son:1998uk,Schafer:1998na,Schafer:1999jg,Hsu:1999mp,Braun:2018bik,Braun:2019aow}, as they also do
in condensed-matter theory (see Refs.~\cite{RevModPhys.66.129,Honerkamp4,Diehl:2009ma,Metzner:2011cw} for reviews). 
Moreover, RG studies provide us with detailed insights into the phase structure and thermodynamics of the theory 
of the strong interaction at lower densities (see, e.g., Refs.~\cite{Braun:2009gm,Strodthoff:2011tz,Strodthoff:2013cua,Morita:2013tu,%
Aoki:2014ola,Khan:2015puu,Jung:2016yxl,Yokota:2016tip,Fu:2016tey,Rennecke:2016tkm,%
Almasi:2017bhq,Fu:2018swz,Braun:2018bik,Fu:2019hdw,Braun:2019aow,Otto:2019zjy,Tripolt:2020irx,Braun:2020ada} 
for some recent advances and Ref.~\cite{Dupuis:2020fhh} for a review) as well as with constraints from quark-gluon dynamics 
for the equation of state of strong-interaction matter over a wide 
range of densities~\cite{Leonhardt:2019fua}.

The central object for a field-theoretical description of quantum systems is the so-called quantum 
effective action. Its computation requires a suitable regularization and renormalization procedure. In general, the 
regularization prescription generates terms depending on a UV cutoff scale~$\Lambda$ which are then absorbed as counter terms in the underlying 
bare action as part of the renormalization procedure. 
With respect to studies of systems at finite density, one may now be worried that the counter terms depend on the chemical potential. 
However, this is not necessarily the case. Indeed, it turns out that the partition function is
invariant under a change of the chemical potential, provided the latter does not exceed a critical value~\cite{Marko:2014hea,Khan:2015puu}.
This is known as the Silver-Blaze property of quantum field theories~\cite{Cohen:2003kd}. 
Assuming that the regularization prescription does not violate the symmetry associated with this invariance, 
it follows that the 
counter terms are indeed independent of the chemical potential~\cite{Marko:2014hea,Khan:2015puu}. 

In RG studies, which are at the heart of the present work, fluctuations are integrated out successively around a given point in momentum space. This procedure implicitly defines 
a regularization and renormalization prescription. 
More specifically, we may choose to integrate out fluctuations from high- to low-momentum scales or around a suitably chosen finite scale, e.g., the 
scale set by the chemical potential~$\mu$ in case of studies of dense matter. Apparently, this choice is delicate since it may induce an explicit breaking of symmetries. 
For example, it may be a natural 
choice to integrate out fluctuations around the Fermi scale~$\mu$ in the presence of a Cooper instability. By this, however, we break explicitly 
the symmetry associated with the aforementioned Silver-Blaze property of quantum field theories, as we shall discuss below. 

These are still very general statements which 
do not yet take into account the fact that an actual computation of the quantum effective action usually involves approximations. 
A prominent and widely used approximation scheme is the so-called derivative expansion, basically corresponding to 
an expansion of correlation functions 
in terms of their external momenta. Such an expansion ultimately requires the specification of an expansion point.
A priori, this point is at our disposal but, loosely speaking, should be chosen such that 
a low-order expansion already allows to capture the most relevant dynamics of the system under consideration. Of course, 
a bias is introduced when we choose a specific expansion point and therefore this has to be considered with great care. Also, the chosen 
expansion point may violate symmetries as the one associated with the Silver-Blaze property. 

The focus of the present work is on issues which may arise in RG studies of dense relativistic matter in the presence 
of a Cooper instability. To this end, we begin our discussion by reviewing 
and extending previous general studies~\cite{Marko:2014hea,Khan:2015puu,Braun:2018svj} of 
the Silver-Blaze property and its consequences from a phenomenological as well as field-theoretical standpoint in Sec.~\ref{subsec:sym}. 
In Sec.~\ref{subsec:derexp}, we then analyze how the Silver-Blaze property and the choice of a specific
expansion point in the derivative expansion 
affects the theoretical predictions for systems where a Cooper instability is expected to govern the underlying dynamics. 
From this analysis, we 
deduce that, in the RG flow, fluctuations should be integrated out around the Fermi surface in order to recover the 
expected Bardeen-Cooper-Schrieffer (BCS) scaling behavior of observables, at least in studies based on a derivative expansion of the 
quantum effective action. Based on these findings, we then construct a new class of regulators for functional RG studies in Sec.~\ref{subsec:rgflowsfs}. 
This class of regulators is suitable to tackle systems  
governed by the presence of a Cooper instability and allows to recover the expected BCS scaling behavior.
Its application is demonstrated with the aid of a simple quark-diquark model 
in Sec.~\ref{subsec:example}. 
Our conclusions including a discussion of the application of the derivative expansion for studies of the theory of the strong 
interaction, i.e., Quantum Chromodynamics (QCD), over a wide range of densities can be found in Sec.~\ref{sec:conc}.

%%%%%%%%%%%%%%%%%%%%%%%%%%%%%%%%%%%%%%%%%%%%%%%%%%%%%%%%
%
\section{Silver-Blaze property}
\label{subsec:sym}
Let us consider a system of fermions and (complex) scalar fields which both couple to a chemical potential~$\mu$. 
This implies that the scalar fields also carry a finite ``fermion number" $F$ (or baryon number~$B$ in case of QCD), i.e., phenomenologically speaking, 
the scalars may be considered as composites of fermions. Examples for such fields are 
diquark fields which carry fermion number~$|F|=2$ (or baryon number~$|B|=2/3$). On the other hand, 
pions composed of a quark and an antiquark carry net fermion number~$F=0$ (baryon number~$B=0$). The sign of~$F$ depends on whether the scalar fields are ``composed" 
of two fermions or two antifermions. Of course, both the fermions and the scalar 
fields may also carry additional quantum numbers, e.g., color in QCD. 
In any case, we shall now consider an action of the following form:
\be
&& S[\bar{\psi},\psi,\phi^{\ast},\phi] \nn\\
&& \quad=  \int_x \Big\{ \bar{\psi} \left( \I \gamma_0(\partial_0  + \mu) + \I\gamma_j\partial_j\right)\psi  \Big. \nn\\[-1mm] 
&& \qquad\qquad \Big. + [(\partial_0 - F\mu)\phi^{\ast}] [(\partial_0 + F\mu)\phi] 
+  (\partial_j\phi^{\ast})(\partial_j\phi)\Big\}  \nn\\[1mm]
&& \qquad\qquad\qquad  
+ V_{\text{int}}[\bar{\psi},\psi,\phi^{\ast},\phi]\,,
\label{eq:action}
\ee
where~$\int_x = \int \diff^4x$ and~$V_{\text{int}}$ describes a set of interactions.
Without loss of generality, we shall assume~$\mu>0$.  
Moreover, we note that we employ Euclidean $\gamma$-matrices with~$\gamma_{\mu}^{\dagger}=\gamma_{\mu}$.

The kinetic terms of the action~$S$ are invariant under the following set of transformations of the fields and 
the chemical potential:
\be
\bar{\psi} \mapsto \bar{\psi}\,{\rm e}^{-\I \alpha x_0}\,,
&\quad& 
\psi \mapsto {\rm e}^{\I \alpha x_0}\psi\,, 
\label{eq:SBTr1}\\
\phi^{\ast} \mapsto \phi^{\ast}\,{\rm e}^{-\I F\alpha x_0}\,,
&\quad& 
\phi \mapsto {\rm e}^{\I F\alpha x_0}\phi\,, 
\label{eq:SBTr2}
%\quad \nn\\
%
\ee
and
\be
\mu \mapsto \mu - \I\alpha \,,
\label{eq:SBTr3}
\ee
where~$\alpha\in {\mathbb{R}}$ is a constant ``angle" and we have promoted~$\mu$ to a 
complex quantity. 
We shall refer to Eqs.~\eqref{eq:SBTr1}-\eqref{eq:SBTr3} as Silver-Blaze transformations. 
Assuming that also the interaction terms contained in~$V_{\text{int}}$ are invariant under this transformation, we conclude that the 
entire action~$S$ is invariant.

Looking at the Silver-Blaze transformation and the action~\eqref{eq:action}, one may be tempted to treat the chemical potential as 
a constant background (gauge) field and introduce corresponding 
``Silver-Blaze-covariant" derivatives of the form~$D_0^{F}=\partial_0+F\mu$ (with~$F\in {\mathbb Z}$). 
Unlike gauge fields, however, we do not integrate over the 
chemical potential in the path integral. The chemical potential is rather an external control parameter.

Let us begin our discussion with the zero-temperature limit and postpone the finite-temperature case to 
the end of this section. The dynamics of the quantum theory associated with the classical action~$S$ is 
determined by the partition function. The latter is related to the following path integral:
\be
\!\!\!\!\!\!{\mathcal Z} = {\mathcal N} \int{\mathcal D}\bar{\psi}{\mathcal D}\psi {\mathcal D}\phi^{\ast} {\mathcal D}\phi
\,{\rm e}^{-S +\int_x {\mathcal J}^{T}\!\cdot\,\varphi}\,,
\label{eq:partsum}
\ee
where~${\mathcal N}$ is a normalization factor.
The vector~${\mathcal J}^{T}=(\bar{\eta},-\eta^{T},J^{\ast},J)$ contains the sources 
for the fermion and scalar fields, respectively, and~$\varphi^{T}=(\psi^{T},\bar{\psi},\phi,\phi^{\ast})$ 
is a collective field vector, only introduced to keep the notation in a compact form.
As the action~$S$ depends on the chemical potential, also the path integral~$\mathcal Z$ depends 
on it.
Evaluating the path integral~$\mathcal Z$ for vanishing sources, 
we obtain the grand-canonical (gc) partition function~${\mathcal Z}_{\text{gc}}$. 

We shall now assume that~$\mathcal Z_{\text{gc}}$ is analytic within some domain of values for the (potentially complex-valued) 
chemical potential. For concreteness, let us assume that $\mathcal Z_{\text{gc}}$ is analytic for~$\Re (\mu) < \muc$. 
Below we shall see that the associated (real-valued) ``critical" value~$\muc$ of the chemical potential is set by 
the (pole) mass and fermion number~$F$ of those fields that couple to the chemical potential. 
From a phenomenological standpoint, this may not come unexpected at all. Indeed, 
the chemical potential is  
the change in free energy when fermions are added to or removed from the system. 
Considering, e.g., a non-interacting system of fermions with mass~$m_{\psi}$ 
for illustration, 
a lower bound for this change in the free energy is given by the mass of the fermions. We therefore expect that 
the fermion density  (strictly speaking, the difference of the density of fermions and antifermions) 
can become finite only for~$\mu \geq m_{\psi}$. For~$\mu < m_{\psi}$, the density 
remains zero and we expect the system to be invariant under a change of the chemical potential. In particular, this 
implies that the partition function~${\mathcal Z}_{\text{gc}}$ does not depend on~$\mu$ for~$\mu < m_{\psi}$. 
This is known as Silver-Blaze property~\cite{Cohen:2003kd}.

Coming back to our general discussion, it follows from the invariance of the action~$S$ and the measure 
of the path integral for the partition function~${\mathcal Z}_{\text{gc}}$ under Silver-Blaze transformations that
\be
{\mathcal Z}_{\text{gc}}\Big|_{\mu} = {\mathcal Z}_{\text{gc}}\Big|_{\mu\to \mu -\I\alpha} \,,
\ee
i.e., the partition function is invariant under a shift of the chemical potential~$\mu$
along the imaginary axis. 
 Assuming that~$\mathcal Z_{\text{gc}}$ is analytic for~$\Re (\mu-\I\alpha) < \muc$, we can perform an analytic continuation 
 of the parameter~$\alpha$, $\alpha\to \I\alpha$ and conclude that~${\mathcal Z}_{\text{gc}}$ does not depend on~$\mu\in {\mathbb R}$ 
 for~$\mu < \muc$. This may be viewed as the mathematical confirmation of 
 our phenomenological line of arguments on the $\mu$-independence of~${\mathcal Z}_{\text{gc}}$ given above. 

Along the lines of Ref.~\cite{Marko:2014hea}, we can now also study the $\mu$-dependence of one-particle-irreducible (1PI) correlation 
functions. To this end, we recall that~$\mathcal Z$ is a functional of the sources and a function of~$\mu$. Using 
again the invariance of the action~$S$ and the measure 
of the path integral under Silver-Blaze transformations, we find
\be
&& \!\!\!\!\! {\mathcal Z}[\bar{\eta},\eta,J^{\ast},J] \Big|_{\mu} \nn\\
&& \; 
= {\mathcal Z}[\bar{\eta}{\rm e}^{-\I\alpha x_0}, {\rm e}^{\I\alpha x_0}\eta, J^{\ast} {\rm e}^{-\I F\alpha x_0}, {\rm e}^{\I F\alpha x_0}J] \Big|_{\mu\to \mu -\I\alpha}
\,.
\ee
Considering now the Legendre transform of~$\ln {\mathcal Z}$ with respect to the 
sources, we eventually arrive at the following 
relation for the quantum effective action~$\Gamma$:
\be
\label{eq:GammaRel}
&& \Gamma[\bar{\Psi}_{\text{cl}}, \Psi_{\text{cl}}, \Phi_{\text{cl}}^{\ast}, \Phi_{\text{cl}}]\Big|_{\mu}\\
&& 
 \;=\! \Gamma[\bar{\Psi}_{\text{cl}}{\rm e}^{-\I\alpha x_0} , {\rm e}^{\I\alpha x_0}\Psi_{\text{cl}}, 
\Phi_{\text{cl}}^{\ast}{\rm e}^{-\I F\alpha x_0}, {\rm e}^{\I F\alpha x_0}\Phi_{\text{cl}}]\Big|_{\mu\to \mu -\I\alpha}\ ,
\nn
\ee
where~$\bar{\Psi}_{\text{cl}}$, ${\Psi}_{\text{cl}}$, $\Phi^{\ast}_{\text{cl}}$, and~$\Phi_{\text{cl}}$ denote the so-called 
classical fields. These fields should {\it not} be confused with those  
which minimize the effective action. 
We can now take functional derivatives of Eq.~\eqref{eq:GammaRel} with 
respect to the classical fields to study the $\mu$-dependence of the $n$-point 1PI correlation functions. 

To illustrate the consequences of Eq.~\eqref{eq:GammaRel} for correlation functions, 
let us take a functional derivative of Eq.~\eqref{eq:GammaRel} with respect 
to~$\bar{\Psi}$ from the left and another one with respect to~$\Psi$ from the right 
and evaluate the resulting expression on the ground state. Assuming that the latter 
is given by vanishing classical fields, we find for the fermionic two-point function in momentum space that
\be
{\Gamma}^{(2)}_{\bar{\psi}\psi}(p_0,\vec{p}^{\,})\Big|_{\mu} = {\Gamma}^{(2)}_{\bar{\psi}\psi}(p_0\!-\!\alpha,\vec{p}^{\,})\Big|_{\mu -\I\alpha}\,.
\ee
Assuming further that~${\Gamma}^{(2)}_{\bar{\psi}\psi}$ is analytic for~$\Re (\mu-\I\alpha) < \muc$,
we can set~$\alpha = -\I\mu$ and find
\be
{\Gamma}^{(2)}_{\bar{\psi}\psi}(p_0,\vec{p}^{\,})\Big|_{\mu} = {\Gamma}^{(2)}_{\bar{\psi}\psi}(p_0\!+\!\I\mu,\vec{p}^{\,})\,.
\label{eq:G11f}
\ee
For the two-point function associated with the complex scalar fields, we 
obtain
\be
{\Gamma}^{(2)}_{\phi^{\ast}\phi}(p_0,\vec{p}^{\,})\Big|_{\mu} = {\Gamma}^{(2)}_{\phi^{\ast}\phi}(p_0\!+\!\I F\mu,\vec{p}^{\,})\,,
\label{eq:G11b}
\ee
provided that~${\Gamma}^{(2)}_{\phi^{\ast}\phi}$ is analytic for~$\Re (\mu-\I F\alpha) < \muc$. 
This analysis can be generalized to all $n$-point functions.  

From this analysis, it follows that, at {\it zero} temperature and~$\mu < \muc$, 
the~$\mu$-dependence of the correlation functions is trivially obtained by simply replacing the 
zeroth components of the four-momenta in the vacuum correlation functions
with suitably $\mu$-shifted zeroth components, see, e.g., Eqs.~\eqref{eq:G11f} and~\eqref{eq:G11b}.
Because of the analytic properties of these functions, it also follows that they 
do {\it not} depend on~$\mu$ for~$\mu < \muc$ at all.

Let us now turn to the critical value~$\muc$ of the chemical potential. From our discussion 
of the two-point functions, we extract that the value of~$\muc$ is set by the pole mass 
and the fermion number~$F$ of those fields which 
couple to the chemical potential. An analytic continuation of the theory 
in the complex~$p_0$-plane is therefore restricted to the domain~$|p_0| < \muc$.
Recall that the pole mass~$m$ of a particle is defined 
by the position of the zero of its inverse propagator for~$\vec{p}=0$:~${\Gamma}^{(2)}(p_0=\I m,0)=0$. 
From our analysis, in particular from Eqs.~\eqref{eq:G11f} and~\eqref{eq:G11b}, we then deduce 
that~$\muc = \min\{m_{\psi}, m_{\phi}/|F|\}$. Here,~$m_{\psi}$ and~$m_{\phi}$ refer to the vacuum 
pole masses of the fermions and bosons, respectively.

Our findings have immediate consequences for the computation 
of the effective action. For example, 
with respect to the derivative expansion 
of the effective action, our analysis suggests 
that the associated expansion
of the correlation functions in external momenta has to be performed 
around the point~$(p_0+\I F\mu,\vec{p}^{\,})=(0,0)$ rather than~$(p_0,\vec{p}^{\,})=(0,0)$ in order to 
preserve the Silver-Blaze property~\cite{Khan:2015puu,Fu:2015naa}. Thus, the expansion point is complex-valued 
for~$\mu >0$. 
An expansion around the point~$(p_0,\vec{p}^{\,})=(0,0)$ breaks explicitly the invariance 
under Silver-Blaze transformations.   
Our discussion implies that a Silver-Blaze-symmetric 
derivative expansion requires to consider different expansion points in the complex $p_0$-plane 
in the underlying expansion of correlation functions. 
In Sec.~\ref{eq:DexpBCS}, we shall discuss issues related to the choice of the expansion point in detail. 

Up to this point, we have ignored that a computation of the effective action requires 
a regularization and renormalization procedure. 
Of course, our conclusions following from the invariance of the theory under Silver-Blaze transformations only hold, 
if this invariance is not violated within that procedure. 

Let us now turn to RG studies of dense relativistic matter. The basic idea of 
RG studies is to integrate out fluctuations successively. For example, in the spirit of Wilson's idea of 
renormalization, we may integrate fluctuations from high- to low-momentum scales. In principle, we may 
also integrate out fluctuations around the Fermi surface. The actual prescription 
specifying the details of the momentum-shell integrations may lead to an explicit breaking 
of the invariance under Silver-Blaze transformations. In the 
following, we shall discuss this aspect by means of the Wetterich equation~\cite{Wetterich:1992yh} which can be directly 
related to Wilson's approach to renormalization on the one-loop level.

The starting point for the derivation of the Wetterich equation~\cite{Wetterich:1992yh} is obtained 
from Eq.~\eqref{eq:partsum} by inserting a regulator term~$\Delta S_k$:
\be
\!\!\!\!\!\!{\mathcal Z}(k) = {\mathcal N} \int{\mathcal D}\bar{\psi}{\mathcal D}\psi {\mathcal D}\phi^{\ast} {\mathcal D}\phi
\,{\rm e}^{-S - \Delta S_k +\int_x {\mathcal J}^{T}\!\cdot\,\varphi}\,.
\label{eq:partsumreg}
\ee
The new term~$\Delta S_k$ is defined as
\be
\!\!\!\!\!\!\! \Delta S_k = \int_x\, \left\{ 
\bar{\psi} {R}_k^{\psi}(\partial_0,\vec{\nabla},\mu)\psi
+
{\phi}^{\ast} {R}_k^{\phi}(\partial_0,\vec{\nabla},\mu) \phi
\right\}\,.
\label{eq:regins}
\ee
This term provides us with a suitable regularization of the path integral. 
In particular, it introduces the additional scale~$k$ which 
screens infrared singularities. Anyhow, for~$k\to 0$, 
we assume that the 
regulator functions appearing in~$\Delta S_k$ vanish, i.e.,~$\lim_{k\to 0}{R}_k^{\psi/\phi}=0$, 
in order to ensure that the original path integral~\eqref{eq:partsum} is recovered in this limit. 
Note that the value of the path 
integral now depends on the scale~$k$ and so do the correlation functions 
derived from it, including the partition function~${\mathcal Z}_{\text{gc}}={\mathcal Z}_{\text{gc}}(k)$.

The regulator insertion~\eqref{eq:regins} is bilinear in the fields. From this, we conclude that it 
does not break the invariance under Silver-Blaze transformations, if the 
momentum-space representation of the regulator functions obeys
\be
R_k^{\psi}(\I p_0,\I \vec{p},\mu) &=& R_k^{\psi}(\I(p_0-\I\mu),\I \vec{p},0)\,,
\label{eq:RkpsiRel}
\ee
and
\be
R_k^{\phi}(\I p_0,\I \vec{p},\mu) &=& R_k^{\phi}(\I (p_0-\I F \mu),\I\vec{p},0)\,,
\label{eq:RkphiRel}
\ee
where it is assumed that~$R_k^{\psi}$ and~$R_k^{\phi}$ are analytic functions of~$p_0$. 
This is in line with the findings from our 
analysis of the $\mu$-dependence of the two-point functions. 

Let us assume for a moment that the regulator functions obey these constraints and that their 
functional form is such that they introduce a mass gap~$m_{\text{gap}}\sim k$ into the theory. By construction, this gap depends 
on the scale~$k$ and 
screens divergences in the limit~$p\to 0$ for~$\mu=0$.
Note that divergences of this type are 
screened by the chemical potential for~$\mu>0$.
In any case, we deduce that the regulator induces a shift of the pole positions of the propagators at finite~$k$.
It is important to add that the regulator should not lower the pole mass of the lowest lying state with finite fermion 
number for~$k>0$ since this potentially leads to a violation of the Silver-Blaze property as well. 
Unfortunately, this appears to be a generic feature of conventionally used momentum cutoffs, see Ref.~\cite{Khan:2015puu} 
for a discussion. 

Since the fermion number~$F$ together with the vacuum pole masses of those fields  
coupled to the chemical potential determine the critical 
value~$\muc$ of 
the chemical potential, the latter also becomes $k$-dependent,~$\muc=\muc(k)$. For values of~$k$ (much) greater 
than any of the pole masses, we even expect~$\muc \sim k$.  
For~$k\to 0$, the $k$-dependent mass gaps in the propagator 
then approach the physical pole masses from above. 
For~$k>0$, it follows by
repeating our analysis for general correlation functions  
that the $k$-dependent correlation functions for finite~$\mu$ are identical to those for~$\mu=0$, provided 
that~$\mu < \muc(k)$. Indeed, for~$\mu < \muc(k)$, the $\mu$-dependence of the $k$-dependent correlation functions is obtained by simply replacing~$p_0$ 
with~$(p_0 + \I F\mu)$ in the corresponding vacuum correlation functions. In particular, 
we find that~${\mathcal Z}_{\text{gc}}(k)$ is constant for~$\mu < \muc(k)$. 
An important consequence of these observations regarding the $\mu$-dependence of the 
correlation functions is that the initial 
conditions for these functions at the scale~$k=\Lambda > c_{\Lambda}\mu$ 
in the RG flow (i.e., the counter terms for the correlation functions in the terminology 
of renormalization theory) are identical to those in the limit~$\mu\to 0$~\cite{Marko:2014hea}. The actual value 
of the constant~$c_{\Lambda}>0$ depends on the details of the functional form of the regulator functions. 
In any case, even if the regulator preserves the Silver-Blaze symmetry, the expansion scheme employed 
to compute the effective action may still break the invariance under Silver-Blaze transformation. For example, 
as discussed above, 
a derivative expansion of the effective action around the point~$(p_0,\vec{p}^{\,})=(0,0)$ breaks the 
invariance under such transformations.

In case of regulators which do not fulfill the constraints~\eqref{eq:RkpsiRel} and~\eqref{eq:RkphiRel}, the invariance under 
Silver-Blaze transformations is explicitly broken and therefore also the initial conditions of the RG flow 
in general depend on the chemical potential. While the dependence of the initial conditions on the 
chemical potential 
becomes parametrically suppressed if the initial scale is chosen sufficiently large,~$\Lambda \gg \mu$, the 
explicit breaking of the Silver-Blaze symmetry by the regulator functions persists.

Let us now analyze the effect of Silver-Blaze transformations on the regularized path integral~\eqref{eq:partsumreg} 
for vanishing source terms, i.e., we consider the $k$-dependent partition function~${\mathcal Z}_{\text{gc}}(k)$. 
Since the action~$S$ and the measure of the corresponding path integral are invariant under 
Silver-Blaze transformations, it suffices to study the variation of the regulator insertion~$\Delta S_k$ 
under an infinitesimal Silver-Blaze transformation. We find
{\allowdisplaybreaks
\begin{widetext}
\be
\Delta S_k[\bar{\psi},\psi,\phi^{\ast},\phi] &\mapsto& \Delta S_k[\bar{\psi},\psi,\phi^{\ast},\phi]
+ \alpha\, \int_p \,\bar{\Psi}(p) \left[ \frac{\partial R_k^{\psi}(\I p_0,\I \vec{p},\mu)}{\partial p_0}  -  \I \frac{\partial R_k^{\psi}(\I p_0,\I \vec{p},\mu)}{\partial \mu}  \right]
 \Psi(p) \nn\\
  && \qquad\qquad\qquad\quad
+ \alpha\, \int_p \,\Phi^{\ast}(p) \left[ F\frac{\partial R_k^{\phi}(\I p_0,\I \vec{p},\mu)}{\partial p_0}  -  \I \frac{\partial R_k^{\phi}(\I p_0,\I \vec{p},\mu)}{\partial \mu}  \right]
 \Phi(p) + {\mathcal O}(\alpha^2)\,,
 \label{eq:DeltaSkVar}
\ee
\end{widetext}
where}~$\int_p = \int \frac{\diff^4p }{(2\pi)^4}$. The Fourier transforms of the fields are defined by
\be
&&\psi (x) = \int_q {\rm e}^{\I q x}\Psi(q)\,,\quad
\bar{\psi} (x) = \int_q {\rm e}^{-\I q x}\bar{\Psi}(q)\,,
\ee
and
\be
&&\phi (x) = \int_q {\rm e}^{\I q x}\Phi(q)\,,\quad
\phi^{\ast} (x) = \int_q {\rm e}^{-\I q x}\Phi^{\ast}(q)\,.
\ee
Note that~$\Psi(q)=\Psi(q_0,\vec{q}^{\,})$ and~$\Phi(q)=\Phi(q_0,\vec{q}^{\,})$. Moreover, 
we have assumed that the regulator functions~$R_k^{\psi/\phi}$ can be expanded in terms of their arguments.
Requiring 
\be
{\mathcal Z}_{\text{gc}}(k)\Big|_{\mu} \stackrel{!}{=} {\mathcal Z}_{\text{gc}}(k)\Big|_{\mu\to \mu -\I\alpha}
\label{eq:ZSBcond}
\ee
for~$\Re(\mu - \I\alpha) < \muc(k)$, we deduce from Eq.~\eqref{eq:DeltaSkVar} that the 1PI two-point functions have 
to obey the following relation for~$\Re(\mu - \I\alpha) < \muc(k)$: 
\begin{widetext}
\be
0&=& \int_p \,\left( \frac{\delta^2 \Gamma_k}{\delta\bar{\Psi}_{\text{cl}} \delta \Psi_{\text{cl}}}\right)^{-1}_0\!\!(p,p)
\left[ \frac{\partial R_k^{\psi}(\I p_0,\I \vec{p},\mu)}{\partial p_0}  - \I \frac{\partial R_k^{\psi}(\I p_0,\I \vec{p},\mu)}{\partial \mu}  \right]
\nn\\
&&  \qquad\qquad\qquad + \int_p \,\left( \frac{\delta^2 \Gamma_k}{\delta \Phi^{\ast}_{\text{cl}} \delta \Phi_{\text{cl}}}\right)^{-1}_0\!\!(p,p)
\left[ F\frac{\partial R_k^{\phi}(\I p_0,\I \vec{p},\mu)}{\partial p_0}  - \I \frac{\partial R_k^{\phi}(\I p_0,\I \vec{p},\mu)}{\partial \mu}  \right]
\,.
 \label{eq:1PISBrel}
\ee
\end{widetext}
The subscript `0' indicates that the second functional derivatives of the effective 
action are evaluated on the minimum of the scale-dependent 
effective action~$\Gamma_k$.  
For example, we have
\be
\left( \frac{\delta^2 \Gamma_k}{\delta \Phi^{\ast}_{\text{cl}} \delta \Phi_{\text{cl}}}\right)^{-1}_0\!\!(p,q) 
= \frac{\delta^2 \ln{\mathcal Z}(k)}{\delta \tilde{J}^{\ast}(p)\delta \tilde{J}(q)}\Bigg|_{\tilde{{\mathcal J}}=0} 
\ee
and similarly for the fermions. Here,~$\tilde{J}$ and~$\tilde{J}^{\ast}$ denote the Fourier transforms of the sources 
for the scalar fields and, correspondingly,~$\tilde{{\mathcal J}}$ is the vector containing the Fourier transforms of all sources. 
For simplicity, we have assumed in Eq.~\eqref{eq:1PISBrel} that the vacuum expectation values of the fields vanish in the minimum 
of the effective action~$\Gamma_k$ for~$\Re(\mu - \I\alpha) < \muc(k)$. In any case, if the identity~\eqref{eq:1PISBrel} is fulfilled 
in the regime defined by~$\Re(\mu - \I\alpha) < \muc(k)$, 
then~$ {\mathcal Z}_{\text{gc}}(k)$ is constant and identical to its value at~$\mu=0$ for $\mu < \muc(k)$. The same holds for all $n$-point 
functions as discussed above.

We immediately deduce from Eq.~\eqref{eq:1PISBrel} that the invariance under Silver-Blaze transformations is preserved if
\be
F\frac{\partial R_k^{\psi/\phi}(\I q_0,\I \vec{q},\mu)}{\partial q_0}  - \I \frac{\partial R_k^{\psi/\phi}(\I q_0,\I \vec{q},\mu)}{\partial \mu} =0\,.
\ee
For example, this equation is fulfilled for the class of regulators implicitly defined by Eqs.~\eqref{eq:RkpsiRel} and~\eqref{eq:RkphiRel}. 

Although Eq.~\eqref{eq:1PISBrel} could potentially be used to derive constraints for the flow of the two-point functions 
to compensate for the regulator-induced violation of the Silver-Blaze symmetry 
in the spirit of Ward identities, we expect this equation to be 
only of limited use for concrete calculations. Indeed, an implementation of such constraints would 
require the knowledge of~$\muc(k)$ which is a dynamically determined quantity. 

Finally, we would like to comment on the case of finite temperature. At finite temperature~$T$ {\it and} chemical potential~$\mu$, the continuous 
symmetry described by Eqs.~\eqref{eq:SBTr1}-\eqref{eq:SBTr3} 
reduces to a {\it discrete} symmetry because of the compactification of the Euclidean time direction:  
\be
\bar{\psi} \mapsto \bar{\psi}\,{\rm e}^{-2\I n\pi T x_0}\,,&\quad&
\psi \mapsto {\rm e}^{2 \I n\pi T x_0}\psi\,,\\
\phi^{\ast} \mapsto \phi^{\ast}\,{\rm e}^{-2\I F n\pi T x_0}\,,&\quad&
\phi \mapsto {\rm e}^{2\I F n\pi T x_0}\phi\,,
\ee
and
\be
\mu \mapsto \mu + 2\I  n \pi T \,,
\ee
with~$n\in {\mathbb{Z}}$. 
At finite temperature, we then have
\be
{\mathcal Z}_{\text{gc}}\Big|_{\mu} = {\mathcal Z}_{\text{gc}}\Big|_{\mu\to \mu + 2\I n\pi T}\,.
\ee
We add that we have~$\mu\to \mu + (2\I n\pi/N)T$ in~$\text{SU}(N)$ gauge theories~\cite{Roberge:1986mm}.
For our analysis, it is now important to realize that the zeroth component of the 
Euclidean four-momentum is discrete at finite temperature. 
Hence, the analytic continuation that underlies our line of arguments in the zero-temperature case cannot be defined uniquely anymore.
From a phenomenological standpoint, this implies that the partition function always exhibits a dependence on the chemical potential 
at finite temperature whereas this is not necessarily the 
case at~$T=0$. 
There, a dependence is only observed if the chemical potential exceeds the critical value~$\muc$ determined by 
the vacuum pole masses and fermion numbers of those fields that are coupled to the chemical potential.

With respect to regulators, we add that the regulator class 
defined by the relations~\eqref{eq:RkpsiRel} and~\eqref{eq:RkphiRel} also respect the discrete symmetry 
present at finite temperature. 
However, the identity~\eqref{eq:1PISBrel} can no longer be applied to control the 
regulator-induced explicit breaking of the Silver-Blaze symmetry since it relies on the consideration of 
infinitesimal transformations. 

We close by noting that, from a more general standpoint, we analyzed the properties 
of $n$-point functions in the complex plane in this section which is also potentially relevant for computations of real-time 
correlation functions~\cite{Floerchinger:2011sc,Haas:2013hpa,Tripolt:2013jra,Tripolt:2014wra,Pawlowski:2015mia,Jung:2016yxl,Yokota:2016tip,Pawlowski:2017gxj,Tripolt:2020irx} 
within the functional RG framework. 

\section{Derivative expansion and BCS scaling}\label{eq:DexpBCS}
\label{subsec:derexp}
With the constraints~\eqref{eq:RkpsiRel} and~\eqref{eq:RkphiRel} for regulators respecting 
the invariance under Silver-Blaze transformations at hand, we now discuss the computation 
of the effective action in a derivative expansion. In the previous section, we have already pointed out 
that, in addition to the regulator, the expansion point associated with the derivative expansion has to be chosen 
carefully in order to ensure that the Silver-Blaze symmetry is not broken explicitly. As we shall now demonstrate, 
however, the expansion point also affects the scaling behavior of physical observables. 
Depending on the chosen expansion point, for example, observables may be found to decrease with increasing 
chemical potential although an increase may be 
expected, e.g., as it is the case in BCS-type models. Moreover, we shall illustrate that, if
Silver-Blaze-symmetric regulators of the type~\eqref{eq:RkpsiRel} and~\eqref{eq:RkphiRel} are used without 
using a Silver-Blaze-symmetric expansion point, loop 
integrals may even turn out to be ill-defined. 

Let us now analyze the derivative expansion of the effective action at finite chemical 
potential at vanishing temperature.
To this end, we employ a simple quark-diquark model with two quark flavors and three colors for illustration. 
Its classical action reads 
\begin{align}
\!\!\! S 
=&\int {\rm d}^4 x\,\Big\{ {\bar{\psi}}\left(\I\partial\!\!\!\slash 
\!+\!\I \mu\gamma_0\right)\psi
 \!+\! \bar{\nu}^2\phi^{\ast}_A\phi_A  \nn\\
 & \; + \I\bar{\psi}\gamma_5\tau_2\phi_{A}^{\ast}T^{A}{\mathcal C}\bar{\psi}^{T}
 - \I\psi^{T}{\mathcal C}\gamma_5\tau_2\phi_{A}T^{A}{\psi}\Big\}\,.
 \label{eq:dqaction}
\end{align}
Here, ${\mathcal C}=\gamma_2\gamma_0$ is the charge conjugation
operator and~$\tau_2$ is the second Pauli matrix living in flavor space. The fermion fields~$\bar{\psi}$ 
and~$\psi$ are understood to contain the two quark flavor degrees of freedom. 
The sum over the color index $A$ runs only over the
antisymmetric color generators $T^{A}$ in the fundamental
representation. 

The complex-valued scalar fields $\phi_A$ carry fermion number~$|F|=2$ since they
represent diquark states of the form~$\phi_{A}\sim ( \bar{\psi}\gamma_5\tau_2 T^{A}{\mathcal
  C}\bar{\psi}^{T})$ with $J^{P}=0^{+}$ for the total angular momentum~$J$ and parity~$P$. 
  The parameter $\bar{\nu}$ can be viewed as an external ``control knob" which 
  can be used to determine the ground-state properties
of this model in the vacuum~($\mu=0$). A general fixed-point
  analysis indeed reveals that two qualitatively distinct ground states are possible, see, e.g., Refs.~\cite{Braun:2017srn,Braun:2018bik,Braun:2019aow} and
  also Ref.~\cite{Alford:1997zt} for a mean-field analysis: 
  first, the ground state in the vacuum limit is already governed by the formation of a diquark condensate, which breaks the
$U_{\rm V}(1)$ symmetry, and, second, the $U_{\rm V}(1)$
symmetry is only broken at finite~$\mu$ because of a 
  Cooper instability in the system but remains intact in the vacuum limit. For the first scenario, we have to choose $\bar{\nu}^2$ to be positive but
sufficiently small. For the second scenario, we have to choose a
sufficiently large value of $\bar{\nu}^2$. Thus, a critical
value $\bar{\nu}_{\ast}$ (associated with a non-Gau\ss ian fixed
point) must exist which separates these two scenarios from each
other.

In the present work, we do not aim at a detailed study of this model. We only employ it to 
demonstrate issues associated with the Silver-Blaze symmetry which arise 
in the computation of the effective action at finite chemical potential. 
To this end, it suffices to compute the effective action of this model in a one-loop approximation where we only take into account purely fermionic loops. 
This would still allow for a finite running of the wavefunction renomalization factors of the diquark fields~\cite{Braun:2018svj}. For 
simplicity, however, we shall drop them as well. 

In the following we shall only consider the RG running of the mass-like parameter~$\bar{\nu}_k^2$. Since~$\bar{\nu}^2_k$ determines 
the curvature of the effective potential at the origin, an analysis of the scale dependence of~$\bar{\nu}^2_k$ allows us to study 
the onset of the spontaneous breakdown of the $U_{\rm V}(1)$ symmetry associated with diquark condensation. 
Indeed, starting the RG flow at a sufficiently large scale~$k=\Lambda \gg \mu$ in the $U_{\rm V}(1)$-symmetric regime, 
we have~$\bar{\nu}^2_k > 0$ at least for a certain range of values of~$k\leq\Lambda$. Depending on the initial value~$\bar{\nu}^2_{\Lambda}\equiv \bar{\nu}^2$, 
the parameter~$\bar{\nu}^2_{k}$ may then change its sign at a scale~$k_{\text{cr}}$, indicating the emergence of 
a nontrivial ground state of the effective potential and therefore of the spontaneous breakdown of the $U_{\rm V}(1)$ symmetry, see, e.g.,  
Refs.~\cite{Braun:2017srn,Braun:2018bik,Braun:2019aow} for a detailed discussion. 
We shall refer to~$k_{\text{cr}}$ as the critical scale. This scale sets the scale for all low-energy quantities~$\mathcal O$ with 
mass dimension~$d_{\mathcal O}$:
\be
{\mathcal O} \sim k_{\text{cr}}^{d_{\mathcal O}}\,.
\label{eq:Okcrd}
\ee
We would like to add that the inverse of~$\bar{\nu}^2_k$ can be traced back to a four-quark interaction 
associated with a diquark channel. In such a formulation, the onset of spontaneous symmetry breaking 
is then indicated by a divergence of the corresponding four-quark coupling~\cite{Braun:2017srn,Braun:2018bik,Braun:2019aow}.

From here on, we assume that the initial condition~$\bar{\nu}^2_{\Lambda}$ for the RG equation of~$\bar{\nu}^2_{k}$ 
has been chosen such that the 
$U_{\rm V}(1)$ symmetry remains intact on all scales for~$\mu=0$. 
For finite~$\mu$, the emergence of a sign change of~$\bar{\nu}^2_k$ 
is then solely due to the presence of a Cooper instability and it was shown~\cite{Son:1998uk,Schafer:1998na} that 
the associated critical scale obeys the following scaling behavior:
\be
k_{\text{cr}} \sim \exp\left(-\frac{c}{\mu^2}\right)\,,
\label{eq:kcrBCS}
\ee
where~$c$ is a dimensionful positive constant which is determined by 
the initial condition~$\bar{\nu}^2_{\Lambda}>0$. 
In the limit~$\mu\to 0$, we have~$k_{\text{cr}}\to 0$, i.e. 
the $U_{\rm V}(1)$ symmetry remains intact on all scales.
Note that the dependence 
of~$k_{\text{cr}}$ on~$\mu$ 
is handed down to physical observables (e.g., the gap) in the IR limit, leading 
to the typical exponential scaling behavior in BCS-type theories. 

Let us now discuss the RG flow of~$\bar{\nu}^2_k$ and the 
resulting scaling of the critical scale~$k_{\text{cr}}$ with~$\mu$ 
in the light of the Silver-Blaze symmetry. For our concrete calculations, 
we employ the Wetterich equation~\cite{Wetterich:1992yh}. The RG flow equation for~$\bar{\nu}^2_k$ 
can then be deduced from the scale-dependent two-point function 
of the complex scalar fields:
\be
\!\!\!\!\!\!\!\! \Gamma^{(2)}_{k,AB}(p,q) := 
\frac{\delta^2 \Gamma_k\left[\bar{\Psi}_{\text{cl}},\Psi_{\text{cl}},\Phi^\ast_{\text{cl}},\Phi_{\text{cl}}\right]}{\delta\Phi^{\ast}_{\text{cl},A}(p)\delta\Phi_{\text{cl},B} (q)}
\Bigg| _{\substack{\bar{\Psi}^{\phantom{\ast}}_{\text{cl}}=\Psi_{\text{cl}}=0 \\ \Phi^\ast_{\text{cl}}=\Phi_{\text{cl}}=0}}\,.
\\[0.01cm]\nn
\ee
For our discussion, it is convenient to separate the trivial color and momentum dependence of this two-point 
function from the rest:
\be
\Gamma^{(2)}_{k,AB}(p,q) = \tilde{\Gamma}^{(2)}_{k}(p)\delta_{AB} (2\pi)^4\delta^{(4)}(p+q)\,.
\ee
The derivative expansion of the effective action can be traced back to an expansion of the $n$-point 1PI 
correlation functions in terms of their external momenta which ultimately requires to specify an expansion point. 
As we do not take into account the running of the wavefunction renormalizations which are obtained by 
taking derivatives of the two-point function with respect to the external momenta, 
the scale-dependence of the parameter~$\bar{\nu}^2_k$ is simply obtained from an evaluation of the  
two-point function 
on the expansion point, i.e., it is associated with the zeroth order of the  derivative expansion. 
\begin{figure}[t]
\centering
\includegraphics[width=0.45\linewidth]{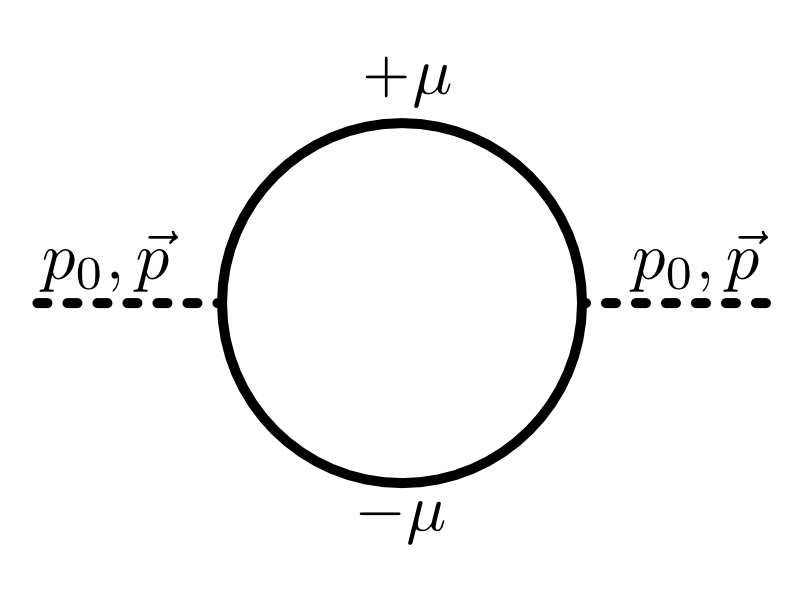}
 \includegraphics[width=0.45\linewidth]{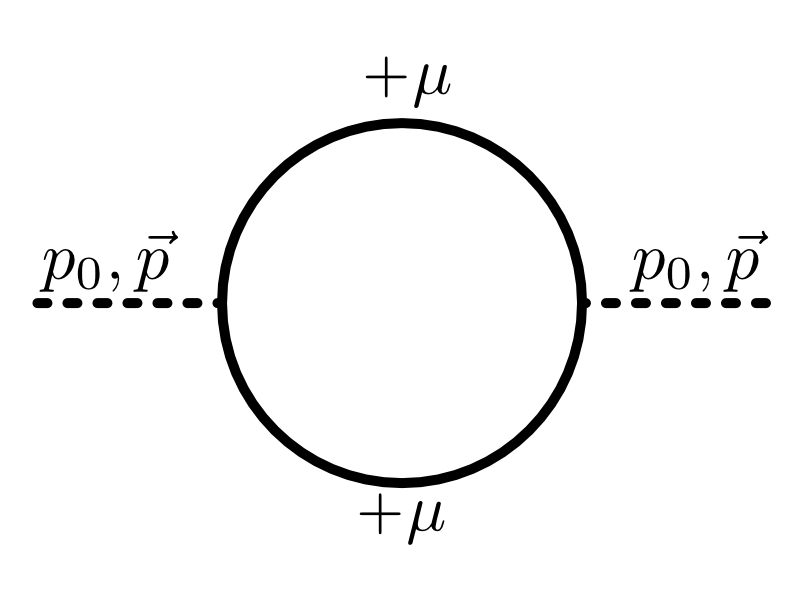}
\caption{1PI diagrams related to the RG flow of the 
two-point function of the complex scalar field. The dashed lines are associated with 
the complex scalar field where~$(p_0,\vec{p}^{\,})$ refers to the external momenta. 
The solid lines are associated with fermions.
}
\label{fig:ld}
\end{figure}

For our computation of the two-point function, we employ a class of regulators 
which only depend on the spatial momenta but neither on the chemical potential nor on the time-like momentum 
and therefore satisfy trivially the Silver-Blaze constraints~\eqref{eq:RkpsiRel} and~\eqref{eq:RkphiRel}. 
Since 
we do not take into account diagrams with internal boson lines in our example study, we only have to regularize fermion lines in the loop diagrams. To be specific, we choose 
the standard form for three-dimensional regulators for fermions~\cite{Braun:2003ii,Schaefer:2004en,Blaizot:2006rj,Litim:2006ag}:
\be
R_k^{\psi} = -\slashed{\vec{p}}\, r(\vec{p}^{\,2}/k^2)\,. 
\label{eq:RpsiStandard}
%\\[-0.2cm]
%\nn
\ee
Here, the so-called regulator shape function~$r$ is to a large extent at our disposal~\cite{Wetterich:1992yh} and will only be specified in 
concrete evaluations of loop diagrams below. 
With this class of regulators, we find the following flow equation for the two-point function of the complex scalar fields: 
\begin{widetext}
\be
\partial_t \tilde{\Gamma}^{(2)}_{k}(p_0,\vec{p}^{\,})
 = -8\int_q \tilde{\partial}_t\ \frac{(p_0 + q_0^{-}) q_0^{+} + (\vec{p}+\vec{q}^{\,})\cdot \vec{q}^{\,} (1+r ((\vec{p}+\vec{q}^{\,})^2/k^2 ) ) (1+r ( \vec{q}^{\,2}/k^2 ) )}{( (p_0 + q_0^{-})^2 + (\vec{p}+\vec{q}^{\,})^2
 (1+r ( (\vec{p}+\vec{q}^{\,})^2/k^2 ) )^2 ) ((q_0^{+})^2 + \vec{q}^{\,2} (1+r ( \vec{q}^{\,2}/k^2 ) )^2 )} \,,
 \label{eq:g2flow}
\ee
\end{widetext}
where $t=\ln(k/\Lambda)$ is 
the RG time with~$\Lambda$ being the initial RG scale,~$\bar{r}_{\vec{q}}=r(\vec{q}^{\,2}/k^2)$,~$q_0^{\pm}=q_0 \pm \I\mu$, and~$\tilde{\partial}_t = (\partial_t r)\frac{\partial}{\partial r}$.
Note that the integral on the right-hand side is associated with the loop integral depicted in Fig.~\ref{fig:ld} (left).

Since we have chosen a class of regulators that respects the Silver-Blaze symmetry and we have also not yet chosen an expansion point for the 
derivative expansion, the RG flow 
of the fully momentum-dependent two-point function does not violate the Silver-Blaze symmetry.
In the spirit of the derivative expansion, we now evaluate the flow equation~\eqref{eq:g2flow} 
on a given expansion point. 
This corresponds to a projection of this flow equation onto the flow equation for the parameter~$\bar{\nu}_k^2$.

We begin by choosing an expansion point which respects the Silver-Blaze symmetry. As discussed in the previous section, this requires to choose 
a point in the complex $p_0$-plane. A possible choice 
is~$(p_0=2\I\mu, \vec{p}=0)$. The RG flow equation for~$\bar{\nu}_k^2$ then reads:
\be
\partial_t \bar{\nu}_k^2 &=&\frac{1}{V_{\text{4}d}}\partial_t \tilde{\Gamma}^{(2)}_{k}(p_0=2\I\mu,\vec{p}=0)\nn\\
&=&
-8\int_q \tilde{\partial}_t\ \frac{1}{(q_0^{+})^2 + \vec{q}^{\,2}(1+r(\vec{q}^{\,2}/k^2))^2}\,,
\label{eq:nu2SBP}
\ee
where~$V_{\text{4}d}=\int {\diff}^4x$. Interpreting the right-hand side of this equation in terms of Feynman diagrams, 
we observe that the integral now corresponds to the loop integral depicted in Fig.~\ref{fig:ld} (right) when evaluated on 
vanishing external four-momentum~$(p_0=0,\vec{p}=0)$. In other words, the evaluation of the original loop integral on 
the Silver-Blaze-symmetric point induces a shift of the chemical potential of one of the internal fermion lines. To analyze the 
consequences of this shift, we evaluate this loop integral for a given regulator. For convenience, we choose the so-called linear regulator 
shape function~\cite{Litim:2006ag},
\be
r_{\text{lin}}(x)=\left(\frac{1}{\sqrt{x}}-1\right)\theta(1-x)\,,
\label{eq:r3dlin} 
\ee
in Eq.~\eqref{eq:nu2SBP} and obtain
\be
\partial_t \bar{\nu}_k^2 = \frac{2 k^2}{3\pi^2}\theta(k-\mu)\,.
\label{eq:dtnu2linreg}
\ee
The solution of this equation reads
\be
 \bar{\nu}_k^2 =  \bar{\nu}_{\Lambda}^2 - \bar{\nu}_{\ast}^2 + \frac{k_{0}^2}{3\pi^2}\,,
 \label{eq:solnu2}
\ee
where~$k_{0}=\max(k,\mu)$ and~$\bar{\nu}_{\ast}^2$ is the aforementioned critical value 
for the parameter~$\bar{\nu}^2 \equiv \bar{\nu}^2_{\Lambda}>0$. 
For the regulator shape function~\eqref{eq:r3dlin}, we find~$\bar{\nu}_{\ast}^2 = \Lambda^2/(3\pi^2)$. 
Choosing~$\bar{\nu}^2_{\Lambda} > \bar{\nu}_{\ast}^2$, we observe that~$\bar{\nu}_k^2$ remains 
positive on all scales and therefore the $U_{\rm V}(1)$ symmetry remains intact even in the IR limit. 
The same is true for~$\bar{\nu}^2_{\Lambda} = \bar{\nu}_{\ast}^2$ and~$\mu=0$ 
but this choice now defines a strongly interacting scale-invariant point which is only destabilized in the presence 
of a finite chemical potential. In particular, we observe that there is no finite critical scale~$k_{\text{cr}}$ 
for~$\bar{\nu}^2_{\Lambda} \geq \bar{\nu}_{\ast}^2$. This implies the absence of dynamical  
 $U_{\rm V}(1)$ symmetry breaking in this case. As a consequence, the scaling of 
 physical observables with the chemical potential~$\mu$ does not agree with the 
 scaling behavior~\eqref{eq:kcrBCS}. We conclude that 
 a derivative expansion around the Silver-Blaze-symmetric point~$(p_0=2\I\mu, \vec{p}=0)$ is not suitable 
 to recover the expected BCS-type scaling behavior of physical observables.  
 
For~$\bar{\nu}^2_{\Lambda} < \bar{\nu}_{\ast}^2$, the curvature~$\bar{\nu}_k^2$ of the effective 
potential may become negative in the RG flow, depending on the actual value of the chemical potential. In fact, 
for~$\mu^2 < \mu_{0}^2 = 3\pi^2(\bar{\nu}_{\ast}^2 - \bar{\nu}^2_{\Lambda})$, we find~$\bar{\nu}_{k=0}^2 < 0$ which indicates 
that the ground state is governed by spontaneous $U_{\rm V}(1)$ symmetry breaking for these values of 
the chemical potential. 
Note that~$\mu_{0}$ defines an {\it upper bound} for the critical value~$\mu_{\text{c}}$ of the chemical potential 
introduced in Sec.~\ref{subsec:sym}. Indeed, the critical scale for~$\mu=0$ is given 
by~$k_{\text{cr}}=\sqrt{3\pi^2(\bar{\nu}_{\ast}^2 - \bar{\nu}^2_{\Lambda})}$ and it follows 
from Eq.~\eqref{eq:solnu2} that this scale does not depend on the chemical potential, provided that 
we choose~$\mu^2 < k_{\text{cr}}^2 = \mu_0^2$. Therefore, all physical observables 
are expected to be independent of~$\mu$ at least for some range of~$\mu < \mu_0$. In any case, since~$k_{\text{cr}}\to 0$
for~$\mu\to\mu_0$, 
the critical scale eventually decreases (rather than increases) when the chemical potential is increased, 
in contradistinction to the case of BCS-type scaling, see Eq.~\eqref{eq:kcrBCS}. 
This is a severe problem even if we would argue that physical (low-energy) observables, such as the gap, 
may still exhibit BCS-type scaling behavior although the critical scale does not. In the present case, however, this would imply
that the low-energy observables increase continuously with increasing chemical potential while the critical scale 
decreases and eventually 
tends to zero above some critical value.

From a phenomenological standpoint, one may argue that the derivative expansion should not be anchored 
at the Silver-Blaze-symmetric point in the complex $p_0$-plane but rather at the conventional point~$(p_0=0,\vec{p}=0)$. 
For the latter point, the flow equation for~$\bar{\nu}_k^2$ is readily obtained from the flow equation~\eqref{eq:g2flow}: 
\be
\partial_t  \bar{\nu}_k^2 \! =\!
-8 \int_q \tilde{\partial}_t\ {\mathcal I}_k(q_0^{+},q_0^{-},\vec{q}^{\,})
\ee
with
\be
&& \!\!\!\! {\mathcal I}_k(q_0^{+},q_0^{-},\vec{q}^{\,}) \nn\\ 
&& \quad = \frac{q_0^{+}q_0^{-} \!+\! \vec{q}^{\,2} (1\!+\! r)^2}
{( ( q_0^{+})^2 \!+ \! \vec{q}^{\,2} (1 \!+\! r)^2) ((q_0^{-})^2 \!+\! \vec{q}^{\,2} (1\!+\! r)^2)}\,.
\ee
In terms of Feynman diagrams, the integral on the right-hand side is now associated with the diagram 
depicted in Fig.~\ref{fig:ld} (left) evaluated on~$(p_0=0,\vec{p}=0)$. 
As shown in Ref.~\cite{Braun:2017srn}, however, a second-order pole at~$k=\mu$ is hidden in this 
loop diagram, rendering the RG flow of the parameter~$\bar{\nu}_k^2$ ill-defined. This is best seen by 
employing again the regulator shape function~\eqref{eq:r3dlin}. The flow equation for~$\bar{\nu}_k^2$ then reads
\be
\partial_t  \bar{\nu}_k^2 =  \frac{k^4}{3 \pi^2} \left\lbrace \frac{1}{\left(k+\mu \right)^2} + \frac{\text{sgn}(k-\mu)}{(k-\mu )^2}\right\rbrace\,,
\label{eq:dtill}
\ee
which reduces to the flow equation~\eqref{eq:dtnu2linreg} in the limit~$\mu\to 0$ as it should be. 
From the flow equation~\eqref{eq:dtill}, we indeed deduce that the RG flow is not well-defined when we insist on integrating out 
fluctuations from~$k=\Lambda$ (high-energy scale) to $k=0$ (low-energy limit). This is also true 
for regulator shape functions other than Eq.~\eqref{eq:r3dlin} 
and can be traced 
back to the fact that the conventionally used class of regulators~\cite{Wetterich:1992yh,Pawlowski:2005xe} defines 
an RG flow from a given high-energy scale down to the low-energy limit. Thus, 
the aforementioned singularity at~$k=\mu$ is always approached from one side.
The actual appearance of this divergence has its origin in the Cooper instability which is of course not pathologic at all. Only the ``treatment" of this divergence 
with the aid of conventional regulator classes is problematic.

One may now be tempted to argue that this singularity is lifted for any (even infinitesimally) finite temperature~$T$~\cite{Braun:2017srn} and, in practice, 
one should  
only consider flows at finite temperature and extrapolate to the zero-temperature limit afterwards. However, this does not cure the 
actual problem of having an ill-defined flow at zero temperature and therefore this idea should be discarded. 
In this respect, we also note that the parameters of models are usually determined at~$T=0$. 

Leaving the finite-temperature case aside, one may argue that the divergence in the flow is 
cured by the presence of a finite diquark gap (i.e., BCS-type gap in general). However, this requires that the gap has already been generated in the 
RG flow at a scale~$k_{\text{cr}} > \mu$, i.e., before the RG flow ``hits" the divergence at~$k=\mu$. Apparently, 
this involves a tuning 
of the parameters of the model. In particular, it excludes to study the case where, e.g., the 
parameters are chosen such that the $U_{\rm V}(1)$ symmetry of the theory remains intact on all scales for~$\mu=0$.
Therefore, also this 
``strategy" of tuning the parameters of the model should be only considered with great care, if at all,  
since the underlying RG flow is still not well-defined. 
We add that, in first-principles studies of QCD, a tuning of parameters, such as~$\bar{\nu}_{\Lambda}^2$, is not 
even possible since~$\bar{\nu}_{\Lambda}^2$ is initially zero and solely generated by quark-gluon interactions in the form of four-quark 
interaction channels, see also Ref.~\cite{Braun:2019aow}.

From our discussion of the flow equation~\eqref{eq:dtill}, we can now deduce how the divergence in the RG flow can be ``cured", namely 
by gapping fluctuations around the Fermi surface with the aid of an artificial gap. The latter should then be successively 
removed towards the 
endpoint of the RG flow. This requires to construct a regulator that integrates out fluctuations around the Fermi surface and thereby introduces a gap 
for fluctuations at the Fermi surface. 
Let us give a qualitative {\it illustration} for such a prescription by means of the flow equation~\eqref{eq:dtill}. 
Formally, the solution of this RG equation 
reads\footnote{Note that, in more elaborate studies,
the right-hand side of the flow equation~\eqref{eq:dtill} 
also depends on other couplings (e.g., Yukawa-type and four-diquark couplings) which hinders a direct integration of the flow equation.}
\be
\bar{\nu}_k^2 - \bar{\nu}_{\Lambda}^2 \sim \int_{\Lambda}^{k} \diff k^{\prime} 
k^{\prime 3} \frac{\text{sgn}(k^{\prime}\!-\! \mu)}{(k^{\prime} \!-\! \mu)^2}\,.
\label{eq:regbyhand}
\ee
Here, we have dropped prefactors and the contribution analytic at~$k=\mu$ as the latter is irrelevant for our qualitative 
discussion at this point. We now employ a principal value prescription to compute this integral which mimics the effect of the  
implementation of a sharp cutoff around the Fermi scale~$\mu$:
\be
\!\!\!\!\!\!\!\!\!\bar{\nu}_{\bar{k}}^2 - \bar{\nu}_{\Lambda}^2 \sim 
-\int_{\mu - \bar{k}}^{0} \diff k^{\prime} 
\frac{k^{\prime 3} }{(k^{\prime} \!-\! \mu )^2} 
\!+\!  
\int_{\Lambda}^{\mu+\bar{k}} \!\!\diff k^{\prime} 
\frac{k^{\prime 3} }{(k^{\prime} \!-\! \mu )^2}\,.
\label{eq:regbyhand2}
\ee
To obtain this equation, we have set~$k=0$ in Eq.~\eqref{eq:regbyhand}. Note 
that IR divergences as they may occur for~\mbox{$\mu=0$} are screened anyhow by 
the presence of the chemical potential. Moreover, we have introduced a new scale~$\bar{k}$ by hand which plays the role of the 
former RG scale~$k$.\footnote{Note that this statement has to be taken with some care since
we do not recover Eq.~\eqref{eq:regbyhand} for~$\bar{k}\to \Lambda$. In fact, there is no simple map between the scales~$k$ and~$\bar{k}$. 
Therefore, we shall restrict ourselves to~$\bar{k} \ll \mu < \Lambda$ which is sufficient for our line of arguments here.}
In the limit~$\bar{k}\to 0$, we then recover Eq.~\eqref{eq:regbyhand} for~$k\to 0$. In any case, performing 
the integration in Eq.~\eqref{eq:regbyhand2}, we find $\bar{\nu}_{\bar{k}}^2 - \bar{\nu}_{\Lambda}^2 \sim \mu^2 \ln \bar{k}$,
where we dropped non-divergent terms for~$\bar{k}\to 0$ as well as prefactors that are irrelevant for our line of arguments. Using 
that~$\bar{\nu}_{\bar{k}}^2=0$ at the critical scale associated with $U_{\rm V}(1)$ symmetry breaking, we find~$\bar{k}_{\text{cr}}\sim \exp (- c/\mu^2)$ 
for the $\mu$-dependence of the critical scale~$\bar{k}_{\text{cr}}$. Here, the constant~$c>0$ is related to the parameter~$\bar{\nu}_{\Lambda}^2$.
Thus, in line with our discussion of Eq.~\eqref{eq:kcrBCS},
we expect the typical BCS-type 
scaling behavior of low-energy observables since the scale for the latter is set by the scale~$\bar{k}_{\text{cr}}$. 

The BCS-type behavior has indeed been revealed in Refs.~\cite{Son:1998uk,Schafer:1998na} by implementing a sharp cutoff around the Fermi scale~$\mu$ to 
integrate out fluctuations around the Fermi surface rather than following the RG flow of the theory from a high-energy scale down to a low-energy scale. 
With respect to the functional RG approach, we have so far only implemented this idea by hand for illustrational purposes. 
In order to do this in a systematic fashion that is readily generalizable to approximations more involved than the one 
considered in this section, 
we construct a suitable class of fermion regulators in the subsequent section.
The class is very general in the sense that it does not require the use of a sharp cutoff 
which often comes at a price of non-locality~\cite{Pawlowski:2005xe} and also ambiguities in the actual computation of RG flow equations, see, e.g., Ref.~\cite{Braun:2014wja}. 
For non-relativistic theories, such a class of regulators has already been constructed within the functional RG framework and successfully employed 
to study a variety of systems, ranging from condensed-matter systems to ultracold 
atomic gases~\cite{Honerkamp4,Birse:2004ha,Diehl:2009ma,Friederich:2010hr,Metzner:2011cw,Braun:2011pp,Roscher:2015xha}, see Ref.~\cite{Pawlowski:2015mlf} 
for a discussion of optimization of RG flows in this context. 
Unfortunately, a naive 
generalization of these regulators to relativistic theories is already hindered by the chiral symmetry. An additional constraint for our regulator construction is set by the fact that 
we would like to ensure that this new class of regulators reduces to the class of conventionally used regulators which regularize divergences in the low-momentum limit. 
Already at this point, we would like to emphasize that it comes at a price to integrate out fluctuations around the Fermi surface. Loosely speaking, the implementation 
of this idea requires to couple the spatial momenta to the chemical potential which then unavoidably leads to a breaking of the Silver-Blaze symmetry. 
However, as discussed above, we require to break this symmetry anyhow in a derivative expansion by choosing 
a suitable expansion point in order to 
recover the correct long-range behavior of physical observables, i.e., BCS-type scaling.

%%%%%%%%%%%%%%%%%%%%%%%%%%%%%%%%%%%%%%%%%%%%%%%%%%%%%%%%
%
\section{RG flows around the Fermi surface}
\label{subsec:rgflowsfs}
\subsection{Chiral fermions}
\label{subsec:chiralfermions}
For our construction of a regulator suitable to deal with fermions in the presence of a Cooper instability, it is convenient to define the following 
two projectors:
\be
P_{\pm}\equiv P_{\pm}(\vec{p}^{\,})
&=&\frac{1}{2\I}\left(\I \gamma_0 \pm \frac{\slashed{\vec{p}}}{|\vec{p}^{\,}|}\right)\gamma_0\,. 
\label{eq:Ppmdef}
\ee
We have
\be
P_{+} + P_{-} &=& \mathbbm{1}\,,\quad
P_{+}P_{-} = P_{-}P_{+} = 0\,,
\label{eq:PpPm1}
\\
P_{+}P_{+} &=& P_{+}\,,\quad
P_{-}P_{-} = P_{-}\,,
\ee
and
\be
P_{-}\gamma_0 &=& \gamma_0 P_{+}\,,\;
P_{+}\gamma_0 = \gamma_0 P_{-}\,,\;
\{ P_{\pm},\gamma_0\} = \gamma_0\,.
\label{eq:PpPm3}
\ee
Note also that~$P_{\pm}(\vec{p}^{\,})=P_{\mp}(-\vec{p}^{\,})$. 
Basically, the operators~$P_{\pm}$ are projection operators on positive and negative energy solutions of the free Dirac equation, 
thus associated with particle and antiparticle states. Similar forms of these 
projectors are often introduced in quantum field theory textbooks~(see, e.g., Refs.~\cite{Bjorken:1965sts,Pokorski:1987ed}) 
and are also employed in hard dense loop studies (see, e.g., Ref.~\cite{Schafer:2003jn}).
Phenomenologically speaking, the fact that 
\be
\tr\, P_{\pm} = 2\,
\label{eq:trPpm}
\ee
reminds us that the free Dirac equation comes with 
 {\it two} solutions with positive and {\it two} solutions with negative energy. 
 
 In Eq.~\eqref{eq:action}, the kinetic term~$S_{\bar{\psi}\psi}$ for the fermions reads
\be
\label{eq:kineticterm}
S_{\bar{\psi}\psi}
=
\int_p \bar{\Psi} \left\{ -( p_0 - \I \mu)\gamma_0 \!-\! \slashed{\vec{p}} \, \right\} \Psi\,,
\ee
where~$\bar{\Psi}$ and~$\Psi$ are the Fourier transforms of the fields~$\bar{\psi}$ and~$\psi$, respectively. 
For convenience, we define the kinetic operator~$T$:
\be
T =  - (p_0 -{\rm i}\mu)\gamma_0 - \slashed{\vec{p}}\,.
\label{eq:defT}
\ee
With the aid of the projectors~$P_{\pm}$, we can decompose~$T$ as follows:
\be
T = C_{-}P_{-}\gamma_0 + C_{+} P_{+}\gamma_0\,,
\label{eq:Tdecomp}
\ee
where
\be
C_{\mp} = - p_0 - \I(-\mu \mp |\vec{p}^{\,}|)\,.
\ee
Using this decomposition, the kinetic term~$S_{\bar{\psi}\psi}$ for the fermions 
can be split into two parts:
\be
\int_p \bar{\Psi} C_{-}P_{-}\gamma_0\Psi 
\,,\quad\text{and}\quad 
\int_p \bar{\Psi} C_{+}P_{+}\gamma_0\Psi\,.
\ee
We observe that these two terms 
are {\it separately} invariant under chiral transformations. Moreover, both terms 
are also  {\it separately} invariant under Silver-Blaze transformations. 
Under charge conjugation, 
the two terms are not invariant. For~$\mu=0$, however, they can still be transformed into 
each other under charge conjugation. Phenomenologically speaking, this is a consequence of the fact that 
the projectors~$P_{\pm}$ decompose the kinetic term~$S_{\bar{\psi}\psi}$ into particle and 
antiparticle contributions. Only~$\int_p\bar{\Psi}T\Psi$ is invariant under charge conjugation for~$\mu=0$. 
Note that the 
symmetry under charge conjugation is broken explicitly by the presence of a finite chemical potential. 

Using the properties~\eqref{eq:PpPm1}-\eqref{eq:PpPm3} of the projectors~$P_{\pm}$, the inverse 
of the operator $T$ -- which basically appears in the computation of loop diagrams -- is readily constructed:
\be
T^{-1} =C_{-}^{-1} P_{+}\gamma_0 + C_{+}^{-1}P_{-}\gamma_0\,.
\ee
The physical meaning of~$C_{\pm}$ becomes apparent when we switch to Minkowski 
spacetime,~$p_0\to -\I p_0$. We then observe that~$T^{-1}$ exhibits poles at~$p_0=\omega_{\pm}$ with
\be
\omega_{\pm}=\pm |\vec{p}^{\,}| -\mu\,.
\ee
The positions of the poles define nothing but the dispersion relation of relativistic (quasi)particles in the 
presence of a chemical potential, i.e., it is essentially the energy of the (anti)fermions relative to the 
Fermi surface located at~$|\vec{p}^{\,}|=\mu$.

We now switch back to Euclidean spacetime and note 
that the inversion of~$T$ has to be considered with care. In fact, whereas~$C_{-}$ is invertible since $C_{-}\neq 0$ for any finite~$\mu$, 
this is not the case for~$C_{+}$. The latter vanishes for~$p_0=0$ and~$|\vec{p}^{\,}|=\mu$ (i.e., at the Fermi surface) and, strictly speaking,~$T$
is therefore not always invertible. At vanishing chemical potential, this singular point corresponds to the case of vanishing four-momentum of which 
is then taken care by, e.g., employing a mass-like regularization scheme. In the presence of a finite chemical potential, however, 
the use of such a scheme does in general not ``cure" the divergence at the Fermi surface and the divergence associated 
with vanishing four-momentum is screened by the chemical potential anyhow.

Let us continue with the construction of 
a class of regulators which is capable of handling the (quasi)particle dispersion relations in a suitable manner. 
Our starting point is Eq.~\eqref{eq:Tdecomp}, i.e., the decomposition of the kinetic operator~$T$ into positive and negative energy 
solutions relative to the Fermi surface. 
The basic idea is now to treat the modes associated with the two terms as two types of modes which are regularized 
differently in the presence of a finite chemical potential\footnote{Note that the chemical potential indeed allows to distinguish the two types
of modes as it breaks the charge conjugation symmetry.} but still regularized in the same way in the limit of vanishing chemical potential.  
The latter requirement ensures that modes with positive and negative energy are treated in the same way for~$\mu=0$ and 
the regularized kinetic term does therefore not break explicitly the charge conjugation symmetry 
in this limit.
In any case, we also require that the regulator does not break the chiral symmetry of the theory under consideration. 
A general regulator for fermions fulfilling these requirements can be written in the following form:
\be
\!\!\!\!\! R_k^{\psi} = - \I(- \mu \!-\! |\vec{p}^{\,}|) r_{-} P_{-}\gamma_0 - \I(-\mu \!+\! |\vec{p}^{\,}|)r_{+} P_{+}\gamma_0\,,
\label{eq:Rdecomp}
\ee 
where~$r_{\pm}$ are again dimensionless regulator shape functions. From here on, we shall assume that these functions are of the following 
form:
\be
r_{\pm} := r(x_{\pm})\,,
\label{eq:shapedef}
\ee
where
\be
x_{\pm} \, k^2 = (-\mu \pm |\vec{p}^{\,}|)^2\,.
\ee
Note that we use the same functional form~$r$ for the regulator of the two modes which ensures that we recover 
the standard form of the flow equation in the absence of a chemical potential (where both types of modes are then treated identically). 
The use of different functional forms would instead imply that the charge conjugation symmetry is 
broken explicitly by the regularization scheme even in the limit~$\mu\to 0$. In any case, the form 
of the regulator proposed in Eq.~\eqref{eq:Rdecomp} already breaks the Silver-Blaze symmetry by construction which, however, 
appears necessary in a derivative expansion of the effective action in order to recover the expected BCS scaling, see  
Sec.~\ref{subsec:derexp}. 

Regarding the shape functions~$r$, we shall in general assume that~$(1+r)\geq 0$ and that they obey
\be
\lim_{x\to 0} \sqrt{x} \, r(x) > 0
\label{eq:rlowx}
\ee
as well as 
\be
\lim_{x\to \infty} r(x) = 0\,.
\label{eq:rlargex}
\ee
For example, the property~\eqref{eq:rlowx} implies that
\be
r(x_{+}) = \frac{k}{|\mu - |\vec{p}^{\,}||}  + \dots\,.
\ee
Loosely speaking, this ensures that, in the RG flow, fluctuations at the Fermi surface are gapped. 
The property~\eqref{eq:rlargex} ensures that the 
regulator vanishes in the limit of~$k\to 0$ for fixed momentum as well as in the limit~$|\vec{p}^{\,}|\to \infty$ for fixed RG scale~$k$. 
Finally, we add that the property~$(1+r)\geq 0$ for any value of~$x$ is required since loop diagrams may 
otherwise be plagued by artificial divergences.

Possible choices for the shape function are given by Eq.~\eqref{eq:r3dlin} and  
\be
r_{\text{exp}}(x) = -1 + \frac{1}{\sqrt{1-{\rm e}^{-x}}}\,,
\label{eq:shapeexp}
\ee
as well as a polynomial version,
\be
\!\!\!\!\!\!\! r_{\text{p}}(x) =  -1 + \frac{1}{\sqrt{1- \left( \sum_{n=0}^{N} \frac{1}{n!}x^n\right)^{-1}}}\,,\; (N > 2)\,.
\label{eq:shapepol}
\ee
Finally, one may also define an ordinary sharp cutoff with the aid of the shape function.\footnote{The functional 
forms are simply adapted from those in Refs.~\cite{Jungnickel:1995fp,Berges:1997eu,Berges:2000ew,Litim:2000ci,Litim:2001fd,Litim:2001up,Pawlowski:2005xe,Litim:2006ag,Blaizot:2006rj,Floerchinger:2011sc} 
introduced for the conventionally used regulator class.} 
Note that 
the general form of the regulator~$R_k^{\psi}$ together with any of the listed shape functions is not necessarily optimized 
in the spirit of~Refs.~\cite{Litim:2000ci,Litim:2001fd,Litim:2001up,Pawlowski:2005xe}. In fact, such an optimization 
of RG flows in the presence of a finite quark chemical potential is beyond the scope of this work. The 
explicit forms of the shape function are presented here only for illustrational purposes.
\begin{figure}[t]
\centering
\includegraphics[width=\linewidth]{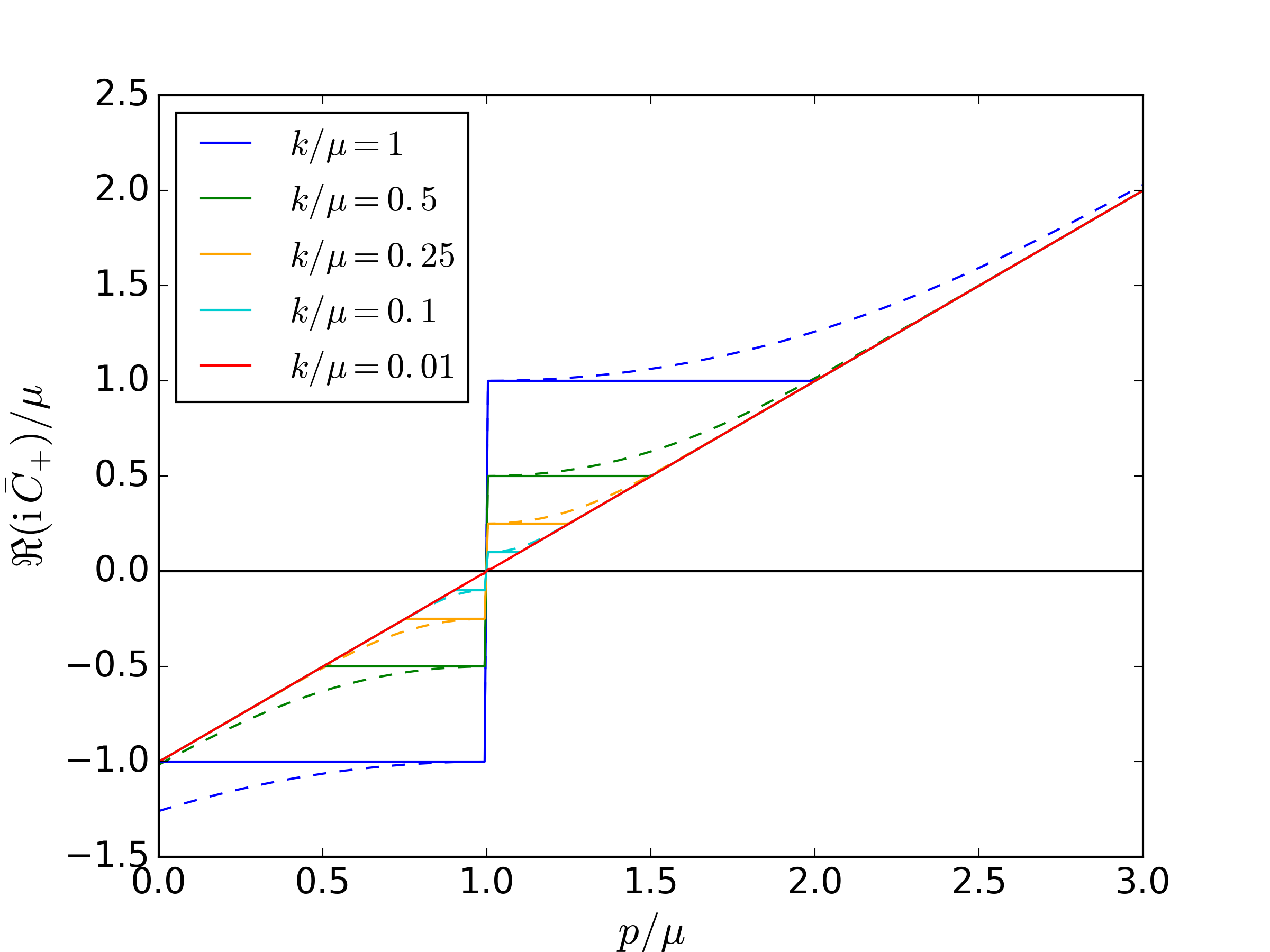}
\caption{$\Re ( \I \bar{C}_{+})/\mu$ 
as a function of~$p/\mu$ for various values of~$k/\mu$ as obtained from employing 
the shape function~\eqref{eq:r3dlin} (solid lines) and 
the shape function~\eqref{eq:shapepol} for~$N=4$ (dashed lines).}
\label{fig:regplot}
\end{figure}

By adding the regulator~$R_k^{\psi}$ to the kinetic operator~$T$, we find
\be
T+R_k^{\psi}  &=& \bar{C}_{-} P_{-}\gamma_0 
+ \bar{C}_{+} P_{+}\gamma_0\,,
\label{eq:regkin}
\ee
where we have introduced regularized ``quasiparticle dispersion relations":
 \be
 \bar{C}_{\mp} = - p_0 - \I(-\mu \mp |\vec{p}^{\,}|)(1+r_{\mp})
 \,.
 \ee
Using the properties~\eqref{eq:PpPm1}-\eqref{eq:PpPm3} of the projectors~$P_{\pm}$, 
the regularized kinetic operator~$T+R_k^{\psi}$ is readily inverted which is now well-defined for all~$p_0$ and~$\vec{p}$.  
The inversion yields the regularized propagator:
\be
(T+R_k^{\psi})^{-1} = \bar{C}_{-}^{-1} P_{+}\gamma_0 
+ \bar{C}_{+}^{-1}P_{-}\gamma_0\,.
\ee 
We note that 
\be
\Re \left( \I \bar{C}_{+}\right) = (|\vec{p}^{\,}|-\mu)(1+r_{+})
\ee
is not positive definite but changes its sign at the Fermi surface. In fact, we have $\Re (\I \bar{C}_{+}) >0$ for~$|\vec{p}^{\,}|>\mu$ 
and $\Re(\I \bar{C}_{+}) <0$ for~$|\vec{p}^{\,}|<\mu$, see also Fig.~\ref{fig:regplot}. Close to the Fermi surface, we find
\be
\Re\left(\I \bar{C}_{+}\right) = k\, \text{sgn}(|\vec{p}^{\,}|-\mu)  + \dots\,,
\label{eq:Cmsignchange}
\ee
implying that~$\Re(\I \bar{C}_{+})$ is discontinuous at the Fermi surface.
In any case, from Eq.~\eqref{eq:Cmsignchange}, we deduce that 
the regulator effectively introduces a gap~$\sim k$ for fluctuations around the Fermi surface. By construction, this 
gap disappears in the limit~$k\to 0$, i.e., in the long-range limit. Hence, fluctuations are integrated out around the Fermi surface. 
For the negative-energy modes associated with~$\bar{C}_{-}$, we note that
\be
\Re\left(\I \bar{C}_{-}\right)  < 0
\ee
even for~$k\to 0$ because of the presence of the chemical potential which, loosely speaking, acts 
as a regulator for these modes.\footnote{One may be tempted to insert only a regulator for the modes 
associated with~$\bar{C}_{+}$ and just let the chemical potential regularize the modes 
with negative energies. However, as indicated above, the vacuum limit ($\mu\to 0$), where both types 
of modes need to be regularized, is then no longer well-defined. A still valid alternative in line with our requirements is 
to consider a regulator of the following form:
\be
R_k^{\psi}= - \I(- \mu \!-\! |\vec{p}^{\,}| r_{-}) P_{-}\gamma_0 - \I(-\mu \!+\! |\vec{p}^{\,}|)r_{+} P_{+}\gamma_0\,,
\nn
%\label{eq:Ralter}
\ee
where the argument of the shape function associated with the negative-energy modes is now chosen to be independent 
of the chemical potential, $r_{-} = r(\vec{p}^{\,2}/k^2)$, but the functional form is still assumed to be 
the same for both types of modes.}

Before we analyze the regularized kinetic operator~$T+R_k^{\psi}$ in the vacuum limit, we would like to give a useful relation for 
the evaluation of the Wetterich equation when our present regularization scheme is employed. It reads
\be
&& \big( T + R_k^{\psi}\big)^{-1} (\partial_t R_k^{\psi}) 
\label{eq:dtRProp}\\[1mm]
&& \quad =  \I(\mu + |\vec{p}^{\,}|) \bar{C}_{-}^{-1} (\partial_t r_{-})P_{+} + \I(\mu - |\vec{p}^{\,}|) \bar{C}_{+}^{-1} (\partial_t r_{+})P_{-}\,. \nn
\ee
We observe that the derivatives of the shape functions, 
which specify the Wilsonian momentum-shell integrations, only appear 
together 
with the corresponding ``propagators"~$\sim\bar{C}_{\pm}^{-1}$. 
With the aid of Eq.~\eqref{eq:trPpm}, the trace over Dirac indices appearing in the Wetterich equation can be 
computed straightforwardly. We find
\be
&&{\text{Tr}_D}\,\left\{ \big( T + R_k^{\psi}\big)^{-1} (\partial_t R_k^{\psi}) \right\} 
\label{eq:TrDWEq}\\[1mm]
&& \quad =  2\I(\mu + |\vec{p}^{\,}|) \bar{C}_{-}^{-1} (\partial_t r_{-})+ 2 \I(\mu - |\vec{p}^{\,}|) \bar{C}_{+}^{-1} (\partial_t r_{+})\,. \nn
\ee
It is also worthwhile to note that the $p_0$-integration can be performed very efficiently with this decomposition using 
Cauchy's residue theorem since the poles in the 
complex $p_0$-plane are readily read off from Eq.~\eqref{eq:TrDWEq}.

With the class of regulators~\eqref{eq:Rdecomp} at hand, we can now in principle 
compute loop diagrams and study the RG flow 
of couplings. Before we illustrate this in Sec.~\ref{subsec:example}, 
we would like to emphasize again that our class of regulators 
does {\it not} violate chiral symmetry and state that, by the proposed decomposition of the regulator 
and the inclusion of the chemical potential, we effectively perform a suitable spectral adjustment of the 
regulator functions. 

Let us finally show that we recover the form of conventionally used 
three-dimensional regulators in the vacuum limit, i.e.,~$\mu\to 0$~\cite{Litim:2006ag,Blaizot:2006rj}.
To this end, we consider the regularized kinetic term~\eqref{eq:regkin} 
for~$\mu=0$.\footnote{Note that~$r_{+}=r_{-}=r$ in this case.}
Looking at the projectors~$P_{\pm}$, we observe that they only depend on the spatial momenta in such a way that 
they are invariant under a rescaling of the spatial momenta with a positive factor:
\be
P_{\pm} (\vec{p}^{\,}(1+r)) = P_{\pm}(\vec{p}^{\,})\,.
\ee
Recall also that~$(1+r)>0$ for~$k>0$. Thus, formally, we can rescale the spatial 
momenta in Eq.~\eqref{eq:regkin} in this way, i.e., we may 
set $\vec{p}_r := \vec{p}^{\,}(1+r)$, and then obtain the following 
expression for~$\mu=0$ by 
employing again the properties of the projectors~$P_{\pm}$:
\be
T+R_k^{\psi} =  -p_0\gamma_0 - \slashed{\vec{p}}_r = -p_0\gamma_0 - \slashed{\vec{p}}(1+r)\,.
\ee
This is nothing but the conventional form of the regularized kinetic operator (for a three-dimensional 
regulator)~\cite{Litim:2006ag,Blaizot:2006rj}. However, we emphasize that, at finite chemical potential, 
our present regulator is different from this conventional choice since it involves the chemical potential and 
therefore treats the modes above and below the Fermi surface in a different way.  

Although our proposed regulator~\eqref{eq:Rdecomp} 
comes with many advantages, it should be stated that it breaks the Silver-Blaze symmetry 
explicitly,\footnote{In fact, the class of regulators defined in Eq.~\eqref{eq:Rdecomp} 
does not obey the condition~\eqref{eq:RkpsiRel}.}
in contrast to conventional three-dimensional regulator functions. However, the latter class of regulators leads 
to ill-defined loop diagrams
because of an insufficient treatment of the Cooper instability, see our discussion in Sec.~\ref{eq:DexpBCS}. 
Moreover, we would like to remind the reader 
that, in a derivative expansion corresponding to an expansion of correlation functions in external momenta 
around a given point in momentum space, we require to break the Silver-Blaze symmetry  
explicitly anyhow by choosing a suitable expansion point such that the 
correct BCS scaling of observables~$\mathcal O$ 
as a function of the chemical potential is recovered,~${\mathcal O}\sim \exp(-c/\mu^2)$ with~$c>0$, 
see Sec.~\ref{eq:DexpBCS}.

%%%%%%%%%%%%%%%%%%%%%%%%%%%%%%%%%%%%%%%%%%%%%%%%%%%%%%%%
%
\subsection{Massive fermions}\label{subsec:massivefermions}
The regulator class constructed in the previous section appears fine for 
chiral fermions. Indeed, it respects 
chiral symmetry and treats modes close to the Fermi surface 
in an adequate manner. However, it is not appropriate to regularize modes with a finite mass close to the Fermi surface. 
It should be noted that, unlike IR divergences in the limit~$\mu\to 0$, divergences at the Fermi surface are not screened 
by a fermion mass term. Therefore, conventional mass-like regulators~\cite{Jungnickel:1995fp,Litim:2006ag,Blaizot:2006rj} 
are not suitable to deal with such divergences appearing at finite~$\mu$.
Indeed, a fermion mass (which may be a parameter or may be generated dynamically by interactions, e.g., by spontaneous chiral symmetry breaking) rather deforms 
the Fermi surface. This deformation effectively leads 
to a shift of the Cooper instability in momentum space. A suitable regulator scheme needs to account for this shift. 

Let us begin our construction of a regulator for massive fermions by enhancing our projectors~$P_{\pm}$:
\be
P_{\pm}(\vec{p}^{\,},m)=\frac{1}{2\I}\left( \I\gamma_0 \pm  \frac{\slashed{\vec{p}}+\I m}{\epsilon}\right)\gamma_0\,,
\ee
where
\be
\epsilon = \sqrt{\vec{p}^{\,2} + m^2}\,.
\ee
From these new projectors we recover the projectors defined in Eq.~\eqref{eq:Ppmdef} in the limit~$m\to 0$.
We again have 
\be
P_{+} + P_{-} &=& \mathbbm{1}\,,\quad 
P_{+}P_{-} =  P_{-}P_{+} = 0\,,\\
P_{+}P_{+} &=& P_{+}\,,\quad
P_{-}P_{-} = P_{-}\,,
\ee
and
\be
\tr\, P_{\pm} = 2\,. 
\ee
For what follows, it is convenient to define a second set of projectors: 
\be
\bar{P}_{\pm}(\vec{p}^{\,},m)=\frac{1}{2\I}\left(\I\gamma_0 \pm \frac{\slashed{\vec{p}}-\I m}{\epsilon}\right)\gamma_0\,,
\ee
Also here, we have
\be
\bar{P}_{+} + \bar{P}_{-} &=& \mathbbm{1}\,,\quad
\bar{P}_{+}\bar{P}_{-} = \bar{P}_{-}\bar{P}_{+} = 0\,,\\
\bar{P}_{+}\bar{P}_{+} &=& \bar{P}_{+}\,,\quad
\bar{P}_{-}\bar{P}_{-} = \bar{P}_{-}\,,
\ee
and
\be
\tr\, \bar{P}_{\pm} = 2\,. 
\ee
Note also that~$\bar{P}_{\pm}(\vec{p},m)=\bar{P}_{\mp}(-\vec{p},-m)$. 
The two sets of projectors are related. First of all, we have~$\bar{P}_{\pm}(\vec{p},0) = P_{\pm}(\vec{p},0)$. Moreover, we find
\be
\bar{P}_{+}\gamma_0 &=& \gamma_0 \bar{P}_{-}\,,\quad
\bar{P}_{-}\gamma_0 = \gamma_0 \bar{P}_{+}\,.
\label{eq:PbP}
\ee
Finally, we note that
\be
\bar{P}_{\pm}^{\dagger}=\bar{P}_{\pm}\,.
\ee
With these operators, we can now construct a regulator for massive fermions which allows us 
to integrate out fluctuations around the Fermi surface. To this end, we first 
add a mass term to the kinetic operator~$T$ defined in Eq.~\eqref{eq:defT}:  
\be
T(m) =  -(p_0 -{\rm i}\mu)\gamma_0 - \slashed{\vec{p}} + \I m\,.
\ee
This operator can be decomposed as follows: 
\be
T(m) = C_{-} \bar{P}_{-}\gamma_0 + C_{+} \bar{P}_{+}\gamma_0\,,
\ee
where 
\be
C_{\mp}\equiv C_{\mp}(m) &=&  - p_0 - \I(-\mu \mp \epsilon)\,.
\ee
The inverse of~$T_m$ reads\footnote{Of course, the inversion is again not well-defined for~$p_0=0$ and~$\epsilon=\mu$. This is taken care 
of below by inserting a suitable regulator.}
\be
T^{-1}(m) =  C_{-}^{-1} P_{+}\gamma_0 + C_{+}^{-1} P_{-}\gamma_0\,.
\ee
Here, we have used Eq.~\eqref{eq:PbP}.

A regulator accounting for the fact that the Fermi surface is effectively deformed 
in the presence of a finite fermion mass is now readily constructed: 
\be
\!\!\!\!\!\!\! R_k^{\psi}(m) =  - \I(-\mu - \epsilon) r_{-} \bar{P}_{-}\gamma_0 - \I(-\mu + \epsilon)r_{+} \bar{P}_{+}\gamma_0\,.
\label{eq:RmDef}
\ee
Examples for possible shape functions~$r_{\pm}=r(x_{\pm})$ are given in Eqs.~\eqref{eq:r3dlin}, \eqref{eq:shapeexp} and \eqref{eq:shapepol}, 
see also Eq.~\eqref{eq:shapedef} and our discussion thereof. However, we now define~$x_{\pm}$ as follows:
\be
x_{\pm} \, k^2  = (-\mu \pm \epsilon)^2\,.
\ee
Note that the mass enters the regulator function. This is required since, unlike ordinary IR divergences 
in the limit~$\mu\to 0$, the Cooper instability is {\it not} cured by the presence of a finite fermion mass. 

Let us now add the regulator~$R_k^{\psi}$ to the kinetic operator~$T(m)$. We find
\be
T(m) +R_k^{\psi} 
 &=& \bar{C}_{-} \bar{P}_{-}\gamma_0 
+ \bar{C}_{+} \bar{P}_{+}\gamma_0\,,
\ee
where
\be
 \bar{C}_{\mp}\equiv \bar{C}_{\mp}(m)  = -p_0 - \I(- \mu \mp \epsilon)(1+r_{\mp})\,.
\ee
Using our relations above, the regularized kinetic operator is readily inverted.  
The inversion yields 
\be
(T(m)+R_k^{\psi})^{-1} = \bar{C}_{-}^{-1} P_{+}\gamma_0 
+ \bar{C}_{+}^{-1}P_{-}\gamma_0\,.
\ee
As discussed above for the massless limit, the Cooper instability is now also regularized for finite fermion masses.
Indeed, close to the Fermi surface, we find for the modes associated with~$\bar{C}_{+}$ that 
\be
\Re\left(\I \bar{C}_{+}\right) = k\, \text{sgn}(\epsilon -\mu)  + \dots\,,
\label{eq:Cmsignchange2} 
\ee
implying that  
the regulator effectively introduces a gap~$\sim k$ for massive fermions around the Fermi surface. 

Note that the statements regarding the decomposition of~$(T+R_k^{\psi})^{-1}(\partial_t R_k^{\psi})$ made in
Eq.~\eqref{eq:dtRProp} for the massless case also apply to the present 
case of finite fermion masses,~$(T(m)+R_k^{\psi})^{-1}(\partial_t R_k^{\psi})$. 

Finally, we would like to add that, even in the limit~$\mu\to 0$, the class 
of regulators defined by Eq.~\eqref{eq:RmDef}  
does not fall into the class of conventionally employed three-dimensional regulators. 
In fact, considering~$\mu=0$ and using\footnote{Recall that~$x_{+}=x_{-}=x=\epsilon^2/k^2$ for~$\mu=0$.}
\be
\bar{P}_{\pm}(\vec{p}^{\,}(1+r),m(1+r)) &=& \bar{P}_{\pm}(\vec{p}^{\,},m)\,,
\ee
we find
\be
T(m) + R_k^{\psi} = -p_0\gamma_0 - (\slashed{\vec{p}}-\I m)(1+r)\,,
\ee
with a regulator shape function depending on~$(\vec{p}^{\, 2} + m^2)/k^2$ rather than only on~$\vec{p}^{\, 2}/k^2$ 
as in the case of conventionally employed three-dimensional regulators. Nevertheless, 
our present class of regulators is also a valid class in the limit~$\mu\to 0$ in the presence of a finite mass. Indeed, it 
rather falls into the class of spectrally adjusted regulators. The latter class  
is often employed in studies of gauge theories, see, e.g., Refs.~\cite{Litim:2002hj,Gies:2002af,Pawlowski:2005xe,Gies:2004hy,Braun:2006jd,Gies:2006wv}. 
In this spirit, we close this section by noting that~$m$ does not have to be a mass in the narrower sense 
of the word. For example,~$m$ may also be related to a chiral (background) field~$\chi$,~$m^2\sim \chi^2$. Moreover, 
if the running of, e.g., wavefunction renormalization factors is taken into account in a given study, 
the inclusion of the latter in our proposed class of
regulators (i.e., also into the shape function)  
may be required in order to ensure that the regulator indeed gaps fluctuations around the Fermi surface.

%%%%%%%%%%%%%%%%%%%%%%%%%%%%%%%%%%%%%%%%%%%%%%%%%%%%%%%% 
%
\section{Example -- Diquark condensation}
\label{subsec:example}  
We now demonstrate the application of the class of regulators defined in Eq.~\eqref{eq:Rdecomp}. To this end, we
consider again the simple quark-diquark model introduced in Eq.~\eqref{eq:dqaction} and compute the flow of the parameter~$\bar{\nu}^2_k$ 
and the four-diquark coupling~$\bar{\lambda}_k$. Although the latter is set to zero in the classical action, 
it is induced in the RG flow by quark-diquark interactions. 
We add that the couplings~$\bar{\nu}_k^2$ and~$\bar{\lambda}_k$ should be viewed 
as the zeroth order terms of an expansion of the two-point and four-point functions in their external momenta, respectively. The expansion point 
is given by~$(p_0,\vec{p}^{\,})=(0,0)$.
For simplicity, we shall only take into account purely fermionic loops in the present example 
study, as also 
done in Sec.~\ref{subsec:derexp}.

The initial condition of the RG flow is given by the classical action~\eqref{eq:dqaction} which is 
invariant under
$U_{\rm V}(1)$ transformations. For the $U_{\rm V}(1)$-symmetric regime, we then find the following set of flow equations:
\be
\partial_t \bar{\nu}^2_k &=& -4\int_p\,\tilde{\partial}_t\left(\frac{1}{p_0^2+(\mu-|\vec{p}^{\,}|)^2(1+r_+)^2}\right.\nn\\
&& \qquad\quad \left.+\frac{1}{p_0^2+(\mu+|\vec{p}^{\,}|)^2(1+r_-)^2}\right)\,,
\label{eq:nu2fe}
\ee
and
\be
\partial_t \bar{\lambda}_k &=& 2\int_p\,\tilde{\partial}_t\left(\left(\frac{1}{p_0^2+(\mu-|\vec{p}^{\,}|)^2(1+r_+)^2}\right)^2\right.\nn\\
&& \quad\; \left.+\left(\frac{1}{p_0^2+(\mu+|\vec{p}^{\,}|)^2(1+r_-)^2}\right)^2\right)\,,
\label{eq:l2fe}
\ee
where~$\tilde{\partial}_t = (\partial_t r_{+})\frac{\partial}{\partial r_{+}} +  (\partial_t r_{-})\frac{\partial}{\partial r_{-}}$. 
We observe that the two flow equations 
are decoupled within our present approximation in the $U_{\rm V}(1)$-symmetric regime.

From the equation for~$\bar{\nu}^2_k$, we can already extract 
the scaling behavior of the critical scale~$k_{\text{cr}}$. Recall that the latter is defined as the scale at which~$\bar{\nu}^2_k$ 
becomes zero,~$\bar{\nu}^2_{k_{\text{cr}}}=0$.
Employing the shape function~\eqref{eq:r3dlin}, we can even compute 
the right-hand side of Eq.~\eqref{eq:nu2fe} analytically. In the limit~$\mu/k \gg 1$,
it then simplifies to~$\partial_t  \bar{\nu}^2_k = 2\mu^2 /\pi^2$. This implies that~$k_{\text{cr}} \sim \exp(-c/\mu^2)$ for~$\mu/\Lambda \ll 1$  
where the $\mu$-independent constant~$c>0$ depends on the initial condition of the flow equation at~$k=\Lambda$. 
In other words, we recover the expected BCS-type scaling behavior of the critical scale, see also Eq.~\eqref{eq:kcrBCS}. 
\begin{figure}[t]
\centering
\includegraphics[width=\linewidth]{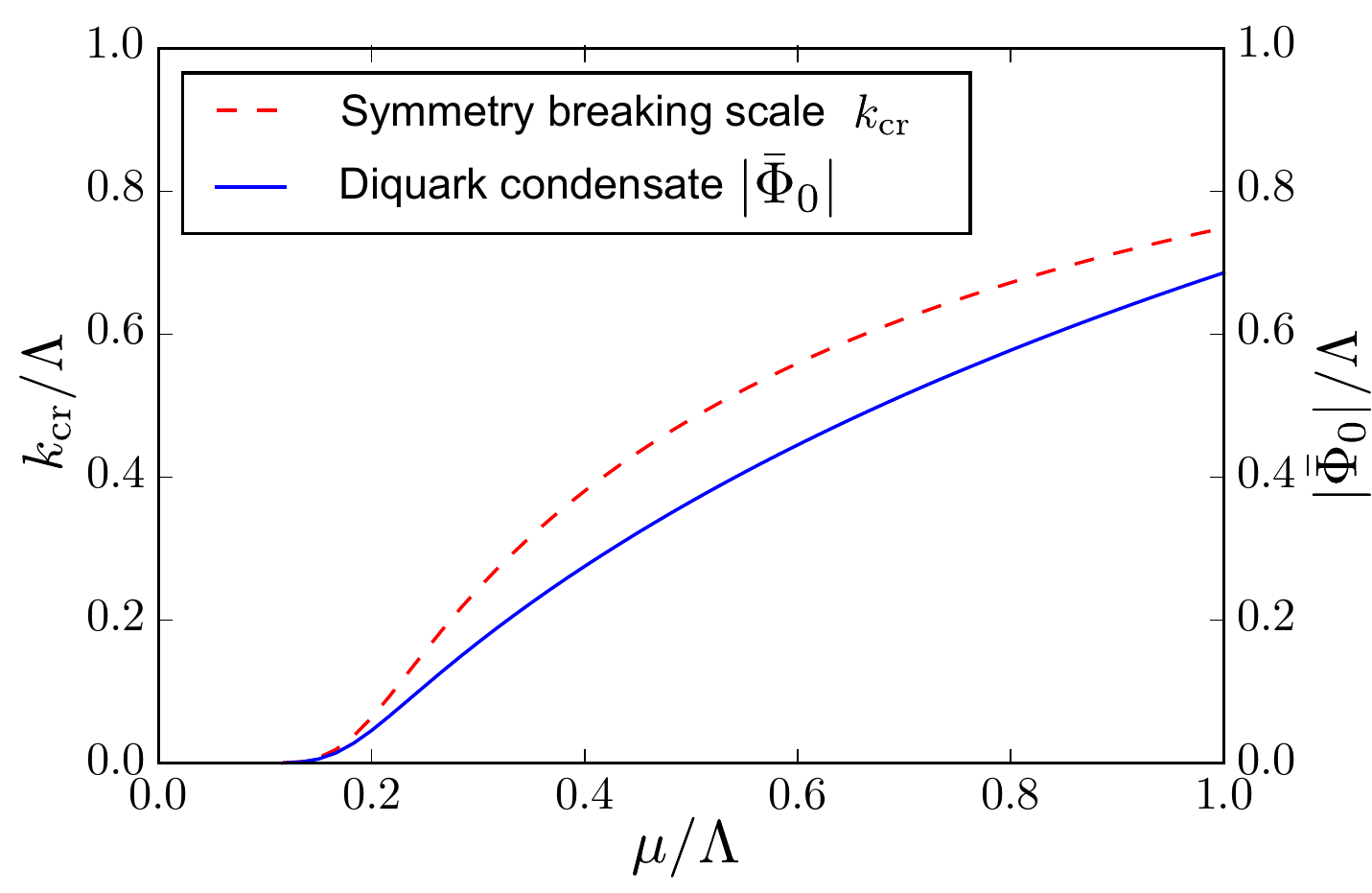}
\caption{Diquark gap~$|\bar{\Phi}_0|$ and critical (symmetry breaking) scale~$k_{\text{cr}}$ in units of the UV scale~$\Lambda$ as a function of 
the dimensionless chemical potential~$\mu/\Lambda$.
}
\label{fig:gap}
\end{figure}

From this analysis of the flow equations in the $U_{\rm V}(1)$-symmetric regime, 
we observe that~$\bar{\nu}^2_k$ changes its sign, indicating the onset of the formation of a diquark condensate. 
For~$k < k_{\text{cr}}$, we then expand the effective action 
around the {\it scale-dependent} ground state~$\bar{\Phi}_0$,~$\Gamma_k \sim \bar{\lambda}_k(|\Phi_{\text{cl}}|^2 - |\bar{\Phi}_0|^2)^2$, 
where~$|\Phi_{\text{cl}}|^2 := \sum_A |\Phi_{\text{cl},A}|^2$ and, for convenience,~$\bar{\Phi}_0$ is chosen to point into the 2-direction of color space 
without loss of generality. 
Using this parametrization of the effective action, 
the flow equations for the low-energy regime read
\be
\partial_t |\bar{\Phi}_0|^2 &=& \frac{2}{\bar{\lambda}_k}\int_p\,\tilde{\partial}_t\left(\frac{1}{p_0^2\!+\!(\mu\!-\! |\vec{p}^{\,}|)^2(1\!+\!r_+)^2+|\bar{\Phi}_0|^2}\right.\nn\\
&& \left.+ \frac{1}{p_0^2+(\mu\!+\!|\vec{p}^{\,}|)^2(1\!+\! r_-)^2+|\bar{\Phi}_0|^2}\right)\,,
\ee
and
\be
\partial_t\bar{\lambda}_k &=&\! 2\int_p \tilde{\partial}_t\left(\left(\frac{1}{p_0^2\!+\! (\mu-|\vec{p}^{\,}|)^2(1\!+\! r_+)^2\!+\!|\bar{\Phi}_0|^2}\right)^2\right.\nn\\
&& \!\!\!\!\! \left.+ \left(\frac{1}{p_0^2\!+\!(\mu+|\vec{p}^{\,}|)^2(1\!+\!r_-)^2\!+\!|\bar{\Phi}_0|^2}\right)^2\right)\,.
\ee
We add that diquark self-couplings of higher order can be taken into account straightforwardly but this is beyond the scope of our 
present example study. 

As an explicit example, we now compute the diquark gap of this model for~$(\bar{\nu}_{\Lambda}/\bar{\nu}_{\ast})^2=4/3$, 
where~$\bar{\nu}_{\ast}^2$ is the value of~$\bar{\nu}_k^2$ at the non-Gau\ss ian fixed point, see also our detailed discussion in Sec.~\ref{subsec:derexp}. 
Phenomenologically speaking, this choice for~$\bar{\nu}^2_{\Lambda}$ implies that the $U_{\rm V}(1)$ 
symmetry is only broken at finite~$\mu$ but remains intact on all scales for~$\mu=0$. In other words, the quarks are ungapped in the vacuum limit.
For our numerical study of the flow equations, we shall employ the
polynomial regulator function~\eqref{eq:shapepol} with~$N=4$ and
set~$\Lambda=0.6\,\text{GeV}$. We then find~$\bar{\nu}^2_{\ast}/ \Lambda^2 \approx 0.065$. 

Our numerical results for the critical scale~$k_{\text{cr}}$ and the value of the diquark gap~$|\bar{\Phi}_0|$ for~$k\to 0$ 
are shown in Fig.~\ref{fig:gap}. 
We observe that~$k_{\text{cr}}$ as well as the diquark gap~$|\bar{\Phi}_0|$ exhibit the same qualitative behavior 
as a function of the chemical potential~$\mu$. This may be viewed as a confirmation of the assumption~$|\bar{\Phi}_0|\sim k_{\text{cr}}$ from 
Sec.~\ref{subsec:derexp}, 
see Eq.~\eqref{eq:Okcrd}. More specifically, for~$\mu\to 0$, we observe 
that~$|\bar{\Phi}_0|$ decreases exponentially whereas it increases monotonically for increasing chemical potential. 
These observations are in line with the expectation that~$\bar{\Phi}_0\sim \exp(-c_{\bar{\phi}}/\mu^2)$ (with~$c_{\bar{\phi}}>0$). 
For a detailed analysis of cutoff effects in this model, we 
refer the reader to Ref.~\cite{Braun:2018svj}. As it is beyond the scope of the present work, we also do not present results for the pressure but 
only refer to App.~\ref{app:td} for a discussion of peculiarities generally appearing in the computation of thermodynamic quantities, such as the 
pressure or the equation of state. 

Overall, we conclude that the class of regulators defined by Eq.~\eqref{eq:Rdecomp} leads to well-behaved RG flows of the 
couplings in the presence of a Cooper instability and the expected BCS-type scaling behavior of physical observables as a function of the chemical 
potential is recovered. Recall that the use of conventionally employed regulator functions leads to ill-defined RG flows, 
see Sec.~\ref{subsec:derexp}, which represented 
the starting point for the construction of the regulator functions employed in this section. 
Nevertheless, a word of caution regarding the regulator class defined by Eq.~\eqref{eq:Rdecomp} is in order: This class is suited
to study systems dominated by a Cooper instability in the long-range limit. In a regime which is governed by the Silver-Blaze property, an expansion 
of the effective action
around a Silver-Blaze-symmetric point should be employed together with a regulator function that does not break the Silver-Blaze symmetry. For example, 
if we use regulators of the form proposed in Sec.~\ref{subsec:rgflowsfs} to compute a diagram of the type as presented in Fig.~\ref{fig:ld} (right) with the external legs being now associated with scalar fields with vanishing  
fermion number~$F$ (e.g., sigma or pion fields), then the diagram does not exhibit the correct scaling behavior as a function of the chemical potential. 
Note that, phenomenologically speaking, this scaling behavior is associated with the screening of interactions in, e.g., the associated 
pseudoscalar channel when then chemical potential is increased. 
From this, it becomes apparent that the expansion point associated with 
a derivative expansion as well as the regularization scheme has to be chosen carefully depending on the 
external parameters, such as the chemical potential. In this sense, a derivative expansion with a given expansion point and regularization scheme should always be viewed 
as an effective description of a theory in terms of the most relevant degrees of freedom and the expansion as well as the scheme should be chosen accordingly. 

\section{Conclusions}
\label{sec:conc} 
In the present work, we discussed issues arising in RG studies of dense systems, with 
a focus on the widely used derivative expansion of the effective action. 
Specifically, 
we critically assessed 
how to apply the derivative expansion to study dense-matter systems governed by a Cooper instability in a systematic fashion, 
including a discussion of the role played by the regularization scheme. 
In particular, we showed that a violation 
of the Silver-Blaze symmetry is required in a derivative expansion of the effective action 
in order to recover the correct BCS scaling behavior of physical observables. 
Based on these formal developments, 
we then introduced a new class of regulators which is suitable to tackle 
dense relativistic matter in the presence of a Cooper instability. 
This class of regulators agrees identically with the standard class of (three-dimensional) regulators in the vacuum limit. 
The latter property is convenient as model parameters 
are often fixed in the vacuum limit by fitting, e.g., vacuum masses or scattering data. It is also 
worth mentioning that this new regulator class does not correspond to simply introducing a sharp cutoff around the Fermi surface. This 
is important as sharp cutoffs are known to lead to ambiguities in the evaluation of loop integrals beyond the lowest order 
of the derivative expansion. 
Finally, we demonstrated the application 
of this class of regulators with the aid of a quark-diquark model and found the typical BCS-type exponential 
scaling behavior of the gap as function of the chemical potential. 

While the use of our new class of regulators is certainly convenient for studies of matter in the presence 
of a Cooper instability, it should also be noted that it introduces an explicit breaking of the Silver-Blaze symmetry and 
is therefore not suited for studies of a regime at small chemical potential where the dynamics is strongly constrained by the 
Silver-Blaze property. This is not a specific problem of the constructed class of regulator functions but 
is in general the case when fluctuations are integrated out around the Fermi surface.
In this sense, the constructed regulator class does not provide ``multi-purpose" regulators.
Our present analysis rather suggests that a consistent and reliable description of the phase diagram and 
thermodynamics of strong-interaction matter based on a derivative expansion in the matter sector is in  
general not possible, {\it if} one insists on employing a derivative expansion anchored at the same point in 
momentum space for all densities. This can be traced back to the fact that a description of QCD 
requires to bridge the gap between a regime governed by chiral symmetry breaking at (very) low densities 
and the presence of a Cooper instability at intermediate and high densities. For
baryon chemical potentials of the order of the mass of the nucleon or below, the dynamics is governed by the Silver-Blaze symmetry. 
Because of the latter, QCD should not exhibit at all a dependence on the baryon chemical 
potential as long as it remains smaller than the nucleon mass. In this regime (also at finite temperature), the derivative 
expansion should be anchored at a Silver-Blaze-symmetric point in momentum space and, within the functional RG approach, regulators respecting 
the Silver-Blaze symmetry should be employed. Note that it is not meaningful to expand the theory around the Fermi surface
since the system does not depend on the chemical potential, at least for sufficiently small chemical potentials at zero temperature.
At high densities, the situation is different. Here, the dynamics 
is expected to be governed by a Cooper instability, we showed that the derivative expansion should not be constructed around 
the Silver-Blaze-symmetric point and also fluctuations in the RG flow should be integrated out around the Fermi surface. The latter 
can be conveniently done with the new class of regulator functions introduced in the present work. Our discussion therefore suggests that 
it is required to switch the approximation scheme as a function of the chemical potential at some {\it a priori} unknown critical chemical potential. Above that point, 
by construction, we violate the Silver-Blaze symmetry in a description of dense strong-interaction matter based on a standard derivative expansion. 
Unfortunately, the regime around this critical chemical potential is most likely of great interest from a phenomenological standpoint as it probably ``accommodates" 
the liquid-gas phase transition and the critical endpoint of the QCD phase diagram. 

This may appear to be only a problem of RG approaches since the RG flow in general covers a wide range 
of scales, ranging from high momentum scales, $\mu/\Lambda \ll 1$, down to the low-momentum regime. 
With respect to other approaches, one may therefore be tempted to argue that the Silver-Blaze 
symmetry is broken anyhow in the infrared limit of QCD at large chemical potentials, as we argued above. 
Unfortunately, the issues discussed in the present work reach beyond RG studies. In fact, as we discussed, the violation 
of the Silver-Blaze symmetry also leaves its imprint in the counter terms in, e.g., conventional loop expansions. 
Indeed, counter terms can only be chosen to be independent of the chemical potential, if the
regularization scheme does not violate the Silver-Blaze symmetry. 

We emphasize that we do not criticize the use of derivative expansions at all since studies resolving the full momentum dependence 
of correlation functions are very costly. We
only would like to highlight issues potentially arising in a description of dense relativistic matter based on 
derivative expansions. At least some of them can be circumvented by carefully choosing the expansion 
point and scheme for a given density regime. With respect to QCD, our present analysis suggests 
that at least two approximation schemes (e.g., in terms of two derivative expansions anchored at two different points) are required to 
reliably describe the properties of strong-interaction matter over a wide range of densities and temperatures, without relying 
on a tuning of, e.g., model parameters. Against this background, we believe that 
our present analysis of derivative expansions
together with our new class of regulator functions for the functional RG approach represents a step forward towards a quantitative 
description of dense strong-interaction matter.

%%%%%%%%%%%%%%%%%%%%%%%%%%%%%%%%%%%%%%%%%%%%%%%%%%%%%%%%
{\it Acknowledgments.--~} 
J.B. is grateful to Marc Leonhardt and Jan M. Pawlowski for many discussion on the subject 
of this work. As member of the {\it fQCD collaboration}~\cite{fQCD}, J.B. and B.S. also would like to 
thank the other members of this collaboration for discussions.
J.B. acknowledges support by the DFG under grant BR~4005/4-1 (Heisenberg program). J.B. 
and S.T. acknowledge support by BMBF under grant 05P20RDFCA.
This work is supported in part by the Deutsche
Forschungsgemeinschaft (DFG, German Research Foundation) -- Projektnummer
279384907 -- SFB 1245.

%%%%%%%%%%%%%%%%%%%%%%%%%%%%%%%%%%%%%%%%%%%%%%%%%%%%%%%%

%
\appendix
\section{Thermodynamics}
\label{app:td}
In this appendix, we would like to employ a non-interacting gas of single-component Dirac fermions to 
highlight peculiarities which arise in the computation of the pressure and are associated with the choice 
of the regularization scheme. 

For a free relativistic Fermi gas at zero temperature, the path integral can be computed analytically:
\be
\!\!\!\Gamma_0 = -\ln {\mathcal Z}_{\text{gc}} = -2 V_{4d} \int_p \ln\left( (p_0 -\I\mu)^2 + \vec{p}^{\,2} \right)\,,
\label{eq:g0fg}
\ee
where~$V_{4d}=\int {\rm d}^4x$,~${\mathcal Z}_{\text{gc}}$ is the grand-canonical partition function, and~$\Gamma_0$ is the effective action 
evaluated at its minimum. Note that the thermodynamic pressure~$p_{T}$
is directly related to the effective action at its minimum~$\Gamma_0$, $p_{T}= - \Gamma_0/V_{4d}$. 
The integral in Eq.~\eqref{eq:g0fg} can be computed 
by suitably inserting an artificial parameter~$m$ into the logarithm~\cite{Bellac:2011kqa}:
\be
\Gamma_0(m) = -2V_{4d} \int_p \ln\left( (p_0 -\I\mu)^2 + \vec{p}^{\,2} +m^2\right)\,,
\ee
where~$m$ acts like a mass term for the fermions. We now take a derivative of this expression with 
respect to~$m$:
\be
\!\!\!\!\!\!\!\!\!\!\! m\frac{\partial}{\partial m}\Gamma_0(m)\! =\! - 4m^2 V_{4d} \int_p \frac{1}{ (p_0 -\I\mu)^2 + \vec{p}^{\,2} +m^2 }\,.
\label{eq:g0cs}
\ee
After performing the integration with respect to~$p_0$ and then with respect to~$m$, we can set~$m=0$ to 
perform the remaining integration with respect to the spatial momenta. We eventually obtain
\be
\frac{1}{V_{4d}}\Gamma_0 = - \frac{\mu^4}{12\pi^2}\,,
\label{eq:g0st}
\ee
where we dropped a divergent constant.

Let us now employ the Wetterich equation to compute the pressure of the free Fermi gas. 
The Wetterich equation for~$\Gamma_0$ reads
\be
\partial_t \Gamma_{0,k} = -\text{tr}_{\rm D}V_{4d}\int_p\,   (\del_t R_k^{\psi})\cdot\left( \Gamma^{(2)}_{\bar{\psi}\psi} + R_k^{\psi} \right)^{-1}.
\label{eq:dtG0}
\ee
Here, the trace has to be taken
with respect to the Dirac indices and
\be
\!\!\!\!\! \Gamma^{(2)}_{\bar{\psi}\psi}(p,q)  = ( -(p_0\!-\!\I\mu)\gamma_0 \!-\! \slashed{\vec{p}}\,)(2\pi)^4\delta^{(4)}(p-q)\,.
\ee
In the present case, 
we can directly integrate the flow equation for~$\Gamma_0$ since the two-point function does not depend on the 
RG scale~$k$. Doing so, we arrive at 
\be
\!\!\!\!\!\!\!\!\! \Gamma_{0} \! -\! \Gamma_{0,\Lambda}  =  -\text{tr}_{\rm D}\int_p\, \left( \ln  \Gamma^{(2)}_{\bar{\psi}\psi}\! -\! \ln \left(\Gamma^{(2)}_{\bar{\psi}\psi} + R_{\Lambda}^{\psi} \right)\right)\,,
\label{eq:G0}
\ee 
where~$\Gamma_{0}\equiv \Gamma_{0,k=0}$. It is worth adding that, even for a scale-dependent two-point function, 
the integrated flow equation for~$\Gamma$ can be written in a form similar to Eq.~\eqref{eq:G0}, see Refs.~\cite{Braun:2007bx,Braun:2010cy} for a discussion. 
In any case,
the first term on the right-hand side of Eq.~\eqref{eq:G0} contains the contribution~$\sim\mu^4$ to the 
pressure. This becomes evident from a comparison with the right-hand side of Eq.~\eqref{eq:g0fg}. The second term contains 
terms depending on the cutoff scale~$\Lambda$ which are divergent in the limit~$\Lambda\to\infty$. 

For an explicit calculation of~$\Gamma_0$, we have to specify the regulator function~$R_k^{\psi}$.  
Let us begin our computation of the pressure by employing the widely used 
class of conventional three-dimensional regulators defined in Eq.~\eqref{eq:RpsiStandard} which do not 
violate the Silver-Blaze symmetry. Employing the shape function~\eqref{eq:r3dlin} in Eq.~\eqref{eq:dtG0}, we find
\be
\!\!\!\!\!\!\! k\frac{\partial}{\partial k} \Gamma_{0,k} \!=\! -4k^2V_{4d} \int_p \theta(k^2\!-\!\vec{p}^{\,2})\frac{1}{  (p_0 -\I\mu)^2 + k^2 }\,.
\label{eq:flatpflow}
\ee
Recall that~$\partial_t = k\partial_k$. From a comparison of Eq.~\eqref{eq:flatpflow} with Eq.~\eqref{eq:g0cs}, we deduce 
that~$k$ has to some extent the effect of a mass gap in case of the conventional class of three-dimensional regulators~\eqref{eq:RpsiStandard}. Note that this is also 
true for the four-dimensional generalization of this regulator class~\cite{Jungnickel:1995fp,Berges:1997eu,Berges:2000ew}. In any case, 
the flow equation for~$\Gamma_{0}$ can be solved analytically. We obtain
\be
\frac{1}{V_{4d}}\Gamma_{0} = \frac{1}{V_{4d}}\Gamma_{0,\Lambda} -\frac{\mu^4}{12\pi^2} + \frac{\Lambda^4}{12\pi^2}\,,
\label{eq:press3dlin}
\ee
which agrees with Eq.~\eqref{eq:g0st} up to an irrelevant constant. 
The latter is cancelled by~$\Gamma_{0,\Lambda}$. 
Thus, we recover the pressure~$p_T=- \Gamma_0/V_{4d}$ of the free relativistic Fermi gas. 

It is worthwhile to add that also sharp cutoffs 
can be implemented in the functional RG approach. 
Following the discussion in Ref.~\cite{Braun:2018svj}, we can obtain a flow equation 
for~$\Gamma_0$ for the (three-dimensional) sharp cutoff: 
\be
\!\!\!\!\!\!\!\!\! k\frac{\partial}{\partial k} \Gamma_{0,k} = \frac{k^3}{\pi^2}V_{4d}\int_{-\infty}^{\infty} \frac{{\rm d}p_0}{2\pi}
\ln ( (p_0 \!-\! \I\mu)^2 \!+\! k^2 )\,.
\label{eq:fsharppflow}
\ee
From this, the thermodynamic pressure can be obtained by integrating over~$k$ from~$k=\infty$ to~$k=0$, relabelling~$k$ in~$|\vec{p}^{\,}|$, and 
then following the line of arguments which led us from Eq.~\eqref{eq:g0fg} to Eq.~\eqref{eq:g0st}. 
Note that a derivation of Eq.~\eqref{eq:fsharppflow} from Eq.~\eqref{eq:dtG0} is delicate as it requires 
the specification of a 
corresponding shape function for the regulator class defined in Eq.~\eqref{eq:RpsiStandard}. For example, one may choose~\cite{Pawlowski:2005xe} 
\be
r_{\text{sharp}} = \frac{1}{\sqrt{\theta(\vec{p}^{\,2} - k^2)}} -1\,.
\ee
Other definitions also exist, see, e.g., Ref.~\cite{Braun:2014wja}. Although the sharp regulator may be appealing at first glance 
since it allows to 
map functional RG flows onto conventional Wilson RG flows 
at one-loop order, it is clear that it should only be used with great care. In fact, 
already from the ambiguity in the definition of the shape function, it appears inevitable that ambiguities arise when this regulator is 
employed, see, e.g., Ref.~\cite{Braun:2014wja} for a concrete example. These ambiguities then also leave their imprint in the predictions 
for the pressure, e.g., in the prefactor of the term~$\sim\mu^4$.

Let us now discuss the flow equation for~$\Gamma_0$ as obtained when our new class of regulators defined by Eq.~\eqref{eq:Rdecomp} is employed.
For illustration purposes, we employ  the shape function~\eqref{eq:r3dlin} which ensures that we recover the 
conventional three-dimensional regulators defined in Eq.~\eqref{eq:RpsiStandard}. This simplifies 
the comparison with our considerations above, in particular with Eq.~\eqref{eq:flatpflow}.
The flow equation  for~$\Gamma_0$ then reads
\be
&&  k\frac{\partial}{\partial k} \Gamma_{0,k} = - 2\I k V_{4d} \int_p \left\{ \theta(k^2 \!-\! (|\vec{p}^{\,}| \!+\! \mu)^2) \frac{1}{-p_0+\I k} \right.\nn\\
 && \;\; \left. + \theta(k^2 \!-\! (|\vec{p}^{\,}| \!-\! \mu)^2) \frac{\text{sgn}(\mu-|\vec{p}^{\,}|)}{-p_0+\I\, \text{sgn}(\mu-|\vec{p}^{\,}|) k}
 \right\}\,.
 \label{eq:fspress}
\ee
First of all, we note that we recover the flow equation~\eqref{eq:flatpflow} for~$\mu=0$. For finite~$\mu$, the two RG flows then differ 
by construction. In fact, the scale~$k$ has no longer the effect of a mass gap but rather of a BCS-type gap. Moreover, the cutoff scales 
for particle and antiparticle excitations are now different and, as discussed 
in Sec.~\ref{subsec:rgflowsfs}, the regulator class defined in Eq.~\eqref{eq:Rdecomp} breaks explicitly the Silver-Blaze symmetry. 
Integrating the flow equation~\eqref{eq:fspress}, we find 
\be
\frac{1}{V_{4d}}\Gamma_{0} =  \frac{1}{V_{4d}}\Gamma_{0,\Lambda} + \frac{\Lambda^4}{12\pi^2} +  \frac{\Lambda^2\mu^2}{2\pi^2} \,,
\label{eq:presssharp}
\ee
The tree-level term~$\Gamma_{0,\Lambda}$ can be computed. For~the presently employed
``Fermi-surface  adapted" regulator~\eqref{eq:Rdecomp} with the shape function~\eqref{eq:r3dlin}, it reads
\be
\frac{1}{V_{4d}}\Gamma_{0,\Lambda} = -\frac{\mu^4}{12\pi^2} -  \frac{\Lambda^4}{12\pi^2} - \frac{\Lambda^2\mu^2}{2\pi^2} + \dots\,,
\label{eq:icfsreg}
\ee
where we dropped terms vanishing in the limit~\mbox{$\Lambda\to\infty$}. 
Thus, we also obtain the correct result for the pressure~\mbox{$p_T=- \Gamma_0/V_{4d}$ from Eq.~\eqref{eq:presssharp}}. 

Note that the presence 
of terms depending on~$\Lambda$ {\it and}~$\mu$ in Eq.~\eqref{eq:presssharp} 
is a generic feature of regulators which integrate out fluctuations 
around the Fermi surface~\cite{Pawlowski:2015mlf}.

Since the regulator class~\eqref{eq:Rdecomp} effectively introduces a gap at the Fermi surface, we close by noting that the Silver-Blaze symmetry 
is similarly broken in conventional mean-field studies of, e.g., quark-diquark models. There, the direct inclusion of a background field with the quantum numbers of the gap  
in the calculation 
leads to a breaking of this symmetry~\cite{Braun:2018svj}. 

\bibliography{qcd}

\end{document}